\documentclass[
aps,
prc,
showpacs,
superscriptaddress,
nofootinbib,
floatfix]
{revtex4}
\usepackage{graphicx,
longtable}
\begin{document}
\title{Neutrino mass hierarchy and electron neutrino oscillation parameters\\ with one hundred thousand reactor events}

%
\author{        F.~Capozzi}
\affiliation{   Dipartimento Interateneo di Fisica ``Michelangelo Merlin,'' 
               Via Amendola 173, 70126 Bari, Italy}%
\affiliation{   Istituto Nazionale di Fisica Nucleare, Sezione di Bari, 
               Via Orabona 4, 70126 Bari, Italy}
\author{        E.~Lisi}
\affiliation{   Istituto Nazionale di Fisica Nucleare, Sezione di Bari, 
               Via Orabona 4, 70126 Bari, Italy}
\author{        A.~Marrone}
\affiliation{   Dipartimento Interateneo di Fisica ``Michelangelo Merlin,'' 
               Via Amendola 173, 70126 Bari, Italy}%
\affiliation{   Istituto Nazionale di Fisica Nucleare, Sezione di Bari, 
               Via Orabona 4, 70126 Bari, Italy}
\begin{abstract}
\vspace*{1cm}
Proposed medium-baseline reactor neutrino experiments offer unprecedented 
opportunities to probe, at the same time, the mass-mixing parameters which govern 
$\nu_e$ oscillations both at short wavelength ($\delta m^2$ and $\theta_{12}$) and at long
wavelength ($\Delta m^2$ and $\theta_{13}$), as well as their tiny interference effects 
related to the mass hierarchy (i.e., the relative sign of $\Delta m^2$ and $\delta m^2$). In order to take full advantage of
these opportunities, precision calculations and refined statistical analyses of event spectra
are required. In such a context, we revisit several
input ingredients, including: nucleon recoil in inverse beta decay and its 
impact on energy reconstruction and resolution, hierarchy and matter effects in the oscillation probability, 
spread of reactor
distances, irreducible backgrounds from geoneutrinos and from far 
reactors, and degeneracies between energy scale and spectrum shape uncertainties.  
We also introduce a continuous parameter $\alpha$, which interpolates smoothly 
between normal hierarchy ($\alpha=+1$) and inverted hierarchy ($\alpha=-1$). 
The determination of the hierarchy is then transformed from a test of hypothesis to 
a parameter estimation, with a sensitivity given by the distance of the true case 
(either $\alpha=+1$ or $\alpha=-1$) from the ``undecidable'' case ($\alpha=0$). 
Numerical experiments are performed for the specific set up envisaged for 
the JUNO project, assuming a realistic sample of $O(10^5)$ reactor events. 
We find a typical sensitivity of $\sim 2\sigma$ to the hierarchy in JUNO, which, however,
can be challenged by energy scale and spectrum shape systematics, whose possible conspiracy
effects are investigated.  The prospective accuracy reachable for the other mass-mixing parameters
is also discussed. 
\end{abstract}
\medskip
\maketitle

\section{Introduction  \label{SecI}}

In $\overline\nu_e$ disappearance searches with reactor neutrinos, 
the survival probability $P_{ee}=P(\overline\nu_e\to\overline\nu_e)$ is generally not
invariant under a swap of the neutrino mass ordering between normal hierarchy (NH) and
inverted hierarchy (IH) \cite{Fo01}. The possible discrimination of the hierarchy via
high-statistics reactor neutrino experiments at medium baseline (few tens of km)  was originally
proposed in \cite{Pe02}
and is now a very active and promising field of research \cite{Work}.
The main idea is to probe,  at the same time, the mass-mixing parameters which govern 
$\nu_e$ oscillations at short wavelength ($\delta m^2, \,\theta_{12}$) and at long
wavelength ($\Delta m^2,\,\theta_{13}$), as well as their tiny interference effects which 
depend on the mass hierarchy, sign$(\Delta m^2/\delta m^2)$  \cite{Pe02,Pe03}. 
The relatively large value of $\theta_{13}$ established in 2012 via 
$\overline\nu_e$ disappearance at short baseline reactors 
\cite{DB01,DB02,RE01,RE02,DC01,DC02}, in agreement with appearance measurements at long-baseline accelerators \cite{T2K1,T2K2,MINO} and with
previous indications from global analyses \cite{Fo11,Fo12}, makes the hierarchy-dependent 
interference effects large enough to be possibly observed in future, dedicated reactor experiments,
such as the so-called RENO-50 \cite{RE50} and JUNO projects \cite{DYB2,DYBR,SanF}.
\footnote{The JUNO (Jiangmen Underground Neutrino Observatory) project \cite{SanF} was 
previously called ``Daya Bay II'' 
\protect\cite{DYB2,DYBR}. 
Within this work, we assume the JUNO project features as reported in \protect\cite{DYB2}.}

The literature in this field is rapidly growing. An incomplete list of pre-2012 studies following \cite{Pe02,Pe03} includes
early tentative experimental projects \cite{HLMA}, theoretical aspects in comparing disappearance probabilities for NH and IH with floating oscillation parameters \cite{Gouv,Nuno,Mina}, prospective data analyses with Fourier transform techniques \cite{Lear,Zha1,Zha2,Baty} also compared with $\chi^2$ analyses  \cite{Ghos}. Post-2012 studies have focused on 
the characterization of more detailed and realistic requirements needed to achieve hierarchy discrimination, such as:  
detector exposure and energy resolution \cite{Gho2,Hagi}, 
peak structure resolution \cite{Ciu0}, 
optimal baselines \cite{Gho2,Hagi}, 
 multiple reactors effects \cite{DYB2,Ciu1,Ciu2}, 
energy scale uncertainties \cite{DYB2,Ghos,Voge,Ciu3,Ciuf}, 
statistical tests of different hierarchy hypotheses   
\cite{Qian,Hagi,Ciu4} and possible synergy \cite{Mi12} with future, independent constraints on 
$\Delta m^2$ \cite{DYB2,Wint,Schw}; see also \cite{Snow} for a very recent review and up-to-date results on prospective data fits. 
All these studies generally find that the hierarchy discrimination
should be possible at a significance level of $\gtrsim 2\sigma$, 
provided that one can achieve unprecedented levels of detector performance and 
collected statistics, which will require the control of several systematics at (sub)percent level. 
Such demanding experimental goals must be matched by accurate theoretical calculations of reactor event spectra 
and by refined statistical analyses. 

In this context, we think it is useful to investigate in more detail
some issues related to the precision calculation and the statistical analysis 
of reactor event spectra, which may provide a useful complement to previous studies in this field.
Whenever possible, we shall highlight analytical results of general 
applicability. Numerical results will instead refer to a specific experimental set-up, namely,
the JUNO configuration described in detail in \cite{DYB2}, including far-reactor and geoneutrino 
contributions.

The structure of our work is as follows. In Sec.~II we present the basic notation and conventions.
In Sec.~III and IV we revisit, from an analytical viewpoint, (sub)percent
spectral effects related to nucleon recoil and to neutrino oscillations in matter, respectively.
We also introduce a useful continuous parameter $\alpha$, which interpolates smoothly 
between normal hierarchy ($\alpha=+1$) and inverted hierarchy ($\alpha=-1$). 
In Sec~V we discuss the ingredients of our numerical and statistical 
analysis of hypothetical samples of $O(10^5)$ reactor events in a JUNO-like experiment. 
In Sec.~VI we present the results of the analysis, and discuss the 
prospective sensitivity to the hierarchy in terms of the distance between the true case 
(either $\alpha=+1$ or $\alpha=-1$) and the ``undecidable'' null case ($\alpha=0$). 
The prospective accuracy expected for the relevant mass-mixing parameters
is also reported. In Sec.~VII we separately discuss several subtle issues
raised by the interplay of energy scale and spectrum shape uncertainties.  
We summarize our work in Sec.~VIII.

\section{Notation}

We present below the basic notation used in this work. Explicit definitions are also needed
to avoid confusion with similar (but not necessarily equivalent) conventions reported in the literature. 

Concerning the neutrino mass-mixing parameters, the squared mass differences and the associated vacuum phases 
are defined as
\begin{equation}
\Delta m^2_{ij} = m^2_i-m^2_j\ ,\ 
\Delta_{ij} = \frac{\Delta m^2_{ij}L}{4E}\ , 
\end{equation}
where $m_i$ are the neutrino masses, $E$ is the neutrino energy, and $L$ is the baseline, in natural units. As in previous papers \cite{Fo01,Fo11,Fo12},
we use a specific notation for the ``small'' and ``large'' squared mass differences (and phases),
\begin{equation}
\delta m^2 = \Delta m^2_{21} > 0\ ,\
\delta =\frac{\delta m^2 L}{4E}>0\ , 
\label{delta}
\end{equation}
\begin{equation}
\Delta m^2 = \frac{1}{2}\left|\Delta m^2_{31}+\Delta m^2_{32}\right|>0\ ,\
\Delta = \frac{\Delta m^2 L}{4E}>0\ .
\end{equation}
Note that, hereafter, $\delta$ will represent the vacuum oscillation phase related
to $\delta m^2$, and not a possible Dirac phase  related to CP violation ($\delta_\mathrm{CP}$). 
The two possible hierarchies are distinguished by a discrete parameter $\alpha$,  
\begin{equation}
\alpha = \left\{ 
\begin{array}{l}
+1\ \mathrm{(normal\ hierarchy)}\ ,\\
-1\ \mathrm{(inverted\ hierarchy)}\ ,
\end{array} \right.
\label{alpha}
\end{equation}
which will be transformed into a continuous variable in Sec.~IV. 
Trigonometric functions of the mixing angles $\theta_{ij}$ (in standard notation \cite{PDGR}) are abbreviated as
\begin{equation}
c_{ij}=\cos\theta_{ij}\ ,\ s_{ij}=\sin\theta_{ij}\ .
\end{equation}

Concerning the inverse beta decay (IBD) process,   
\begin{equation}
\overline\nu_e + p \to e^+ + n\ ,
\end{equation}
the relevant information is contained in the IBD event spectrum $S$ as a function of 
the observed ``visible'' energy of the event. The spectrum $S$ is obtained by integrating out the (unobservable) 
true energies of the incoming neutrino and of the outcoming positron, 
\begin{equation}
S(E_\mathrm{vis}) = \varepsilon(E_\mathrm{vis})\int_{m_e}^\infty dE_e \int_{E_T}^\infty dE 
\left(\sum_i \;{\cal N}_i \;\Phi_i(E) P_i(E)\right) \frac{d\sigma(E,\,E_e)}{dE_e}\;
r(E_e+m_e,\,E_\mathrm{vis})\ ,
\label{spectrum}
\end{equation}
where
\begin{eqnarray}
S(E_\mathrm{vis})				&=& \mathrm{spectrum\ of\ events \ per\ unit\ of\ energy},\\
E 				&=& \overline\nu_e\ \mathrm{energy},\\
E_T				&=& E\ \mathrm{threshold\ for\ IBD},\\
E_e 				&=& \mathrm{true\ positron\ energy\ (total)},\\ 
m_e				&=& \mathrm{positron\ mass}\\
{d\sigma(E,\,E_e)}/{dE_e} &=& \mathrm{IBD\ differential\  cross\ section},\\
E_e+m_e 			&=& \mathrm{true\ visible\ energy\ of\ the\ event},\\
E_\mathrm{vis}	&=& \mathrm{observed\ visible\ energy\ of\ the\ event},\\
r(E_e+m_e,\,E_\mathrm{vis}) &=& \mathrm{energy\ resolution\ function},\\
\varepsilon(E_\mathrm{vis}) &=& \mathrm{detector\ efficiency},\\
	i			&=& \overline\nu_e\ \mathrm{source\ index},\\
\Phi_i(E)		&=& \overline \nu_e\ \mathrm{flux\ (per\ unit\ of\ energy,\ area\ and\ time)},\\ 
P_i(E)			&=& \overline\nu_e\ \mathrm{survival\ probability},\\
{\cal N}_i		&=& \mathrm{normalization\ and\ conversion\ factor}.
\end{eqnarray}
In the above equations, integration over time is implicit: the source fluxes $\Phi_i$ 
or the detector efficiency $\varepsilon$ should be understood either as constants or as time averages, unless otherwise stated.
Further details on these and related ingredients of the analysis are described in the following sections.

\section{Recoil effects in IBD}

The kinematics and dynamics of IBD cross section have been thoroughly studied in \cite{Vog1,Vog2,Viss}. Here we 
revisit nucleon recoil effects on reactor spectra, which are not entirely negligible (as it is often assumed) in the context of 
high-precision experiments. We show that such effects can be included in the calculation of 
(un)binned reactor neutrino event spectra, through appropriate modifications of the energy resolution function.

\subsection{Positron energy spectrum at fixed neutrino energy $E$}

The IBD kinematical threshold is given by 
\begin{equation}
E\geq E_T = \left[(m_n+m_e)^2-m_p^2\right]/2m_p = 1.806\ \mathrm{MeV}\ ,
\end{equation}
where $m_p$ and $m_n$ are the proton and neutron masses, respectively. In the popular ``recoilless'' 
approximation, the positron energy $E_e$ is directly linked to the neutrino energy $E$ via $E-E_e \simeq \Delta_{np}$ (where $\Delta_{np}=m_n-m_p=1.293~\mathrm{MeV})$. 
However, since a small fraction of energy [of $O(E/m_p)$] is carried by the recoiling nucleon, this estimate  provides only an approximate upper bound to $E_e$. More precisely, $E_e$ falls within a well-defined kinematical range, 
\begin{equation}
E_e\in [E_1,\, E_2] \ ,
\end{equation}
where explicit expressions for $E_{1,2}$ can be found, e.g., in \cite{Viss}. For $E$ largely above threshold, 
the boundaries of the neutrino-positron energy difference $E-E_e$ are approximately given by 
\begin{eqnarray}
E-E_2 &\simeq & \Delta_{np} \ ,\\  
E-E_1 &\simeq & \Delta_{np} + 2(E-\Delta_{np})E/m_p\ .
\end{eqnarray}

\begin{figure}[th]
\vspace*{.0cm}
\hspace*{.0cm}
\includegraphics[scale=0.45]{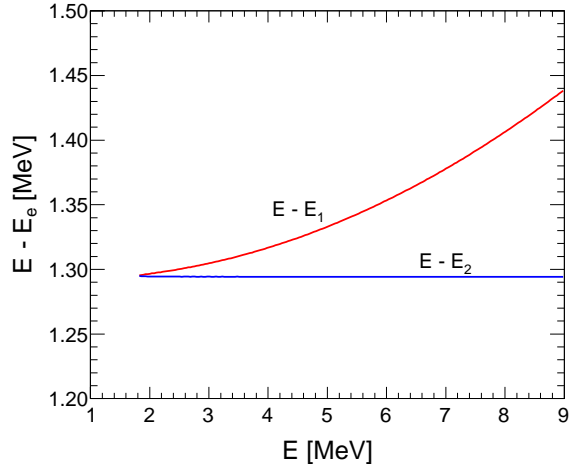}
\vspace*{0.0cm}
\caption{\label{f01}
Inverse beta decay: Range of the difference between the $\overline\nu_e$ energy ($E$) and the $e^+$ energy ($E_e$), as a function of $E$. The extrema are indicated as $E-E_1$ and $E-E_2$. See the text for details.
}
\end{figure}

Figure~1 reports the exact boundaries  (with no approximation) as a function of $E$. From this figure and from the above expressions
it appears that, in the high-energy tail of the reactor spectrum ($E\simeq 6$--8~MeV), recoil corrections can reach the percent level,
comparable to the prospective energy scale accuracy and resolution width \cite{Voge,DYB2} in the same range. We emphasize that
the correction to the recoilless approximation is twofold: at any given $E$, the typical $E_e$ energy is 
displaced at $O(E/m_p)$ and it also acquires a spread of $O(E/m_p)$. Both effects can be taken into account as follows.

Within the narrow range $[E_1,\,E_2]$, the IBD dynamics governs the spectral distribution of $E_e$, i.e., the normalized differential cross section $\sigma^{-1}d\sigma/d E_e$.  Figure~2 shows this distribution in terms of deviations of $E_e$ from its mid-value, $\Delta E_e = E_e-(E_1+E_2)/2$, for selected values of the neutrino energy $E$. For definiteness, we have used the cross section as taken from \cite{Viss}. At small energies, the distributions in Fig.~2 approach the ``Dirac deltas'' expected in the recoilless approximation, while at high energies there is a noticeable spread. For our purposes (see the next subsection) each distribution can be approximated by a ``top hat'' function for $E_e \in [E_1,\,E_2]$:
\begin{equation}
\frac{1}{\sigma(E)}\,\frac{d\sigma(E,E_e)}{d E_e}\simeq \frac{1}{E_2-E_1}\ , 
\label{bettersigma}
\end{equation}
where $\sigma(E)=\int d{E_e}(d\sigma/d E_e)$. We have verified that further corrections related to the slight slopes
in Fig.~2 are completely negligible in the calculation of observable event spectra.

\begin{figure}[bh]
\vspace*{.0cm}
\hspace*{.0cm}
\includegraphics[scale=0.45]{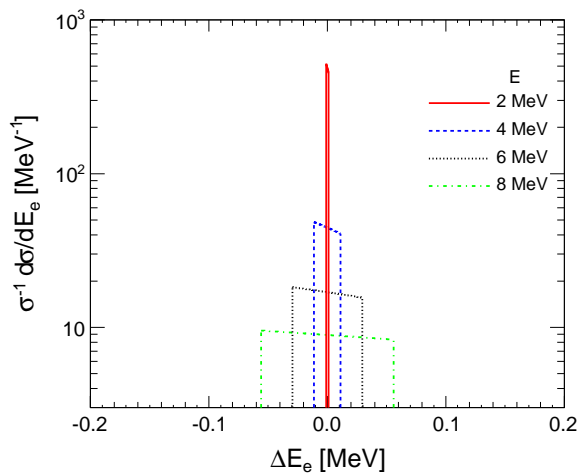}
\vspace*{0.0cm}
\caption{\label{f02}
Inverse beta decay: shape of the $e^+$ energy spectrum for representative values of the $\overline\nu_e$ energy $E$. The spectra are aligned to 
their median value for graphical convenience. 
}
\end{figure}

\subsection{Recoil effects in unbinned spectra}

For any detected IBD event, the observed visible energy $E_\mathrm{vis}$ may differ from the true visible energy $E_e+m_e$, due to intrinsic fluctuations in the collected photon statistics and other possible uncertainties. We assume a gaussian form for the corresponding energy resolution function $r$,
\begin{equation}
r(E_e+m_e,\,E_\mathrm{vis}) = \frac{1}{\sigma_e(E_e)\sqrt{2\pi}}\exp \left[-\frac{1}{2}\left(
\frac{E_\mathrm{vis}-E_e-m_e}{\sigma_e(E_e)}
\right)^2\right] \ ,
\label{gaussian}
\end{equation}
with a prospective width \cite{DYB2}
\begin{equation}
\frac{\sigma_e(E_e)}{E_e+m_e} = \frac{3\times 10^{-2}}{\sqrt{(E_e+m_e)/\mathrm{MeV}}}
\label{width}
\end{equation}
which decreases from $\sim 3\%$ at $E\sim 2$~MeV to $\sim 1\%$ at $E \sim 8$~MeV.

Various assumptions and empirical parametrizations for the width $\sigma_e$ have been studies elsewhere (see, e.g., \cite{DYB2,Hagi,Voge,Gho2,Ciu0,Snow} for recent examples), showing that it is imperative to have $\sigma_e$ as small as possible, i.e., close to the ideal limit of full light collection. The empirical form of $\sigma_e$ in a real detector is actually determined by a combination of calibration experiments and light-yield MonteCarlo simulations, which fix at the same  time the energy scale and the energy resolution, as well as their correlated uncertainties \cite{Tesi}. In this work we do not deal with these subtle experimental aspects, and simply assume $\sigma_e$ as in the above equation; we instead focus on the inclusion of recoil effects of $O(E/m_p)$ which, as noted, can be as large as $O(\sigma_e/E)$ for $E\simeq 8$~MeV.

\begin{figure}[t]
\vspace*{.0cm}
\hspace*{.0cm}
\includegraphics[scale=0.48]{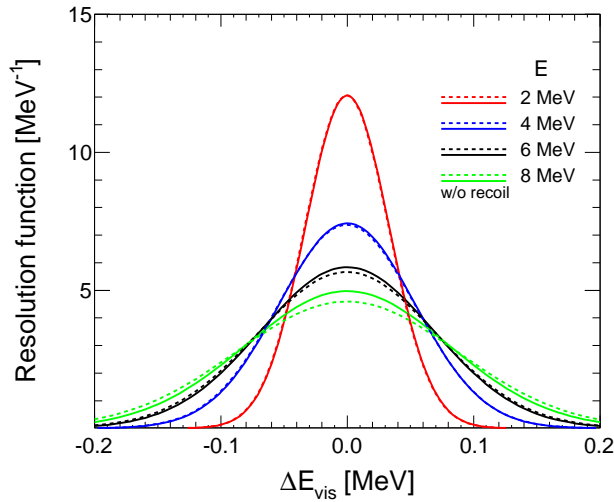}
\vspace*{0.0cm}
\caption{\label{f03}
Energy resolution function without (solid) and with (dashed) the inclusion of nucleon recoil effects, for the same representative
values of the neutrino energy $E$ as in Fig.~2. The functions are aligned to their median value for graphical convenience.
}
\end{figure}

In the approximation of Eq.~(\ref{bettersigma}) and for a gaussian resolution function as in Eq.~(\ref{gaussian}), the inner integral of
the continuous (unbinned) spectrum $S$ in Eq.~(\ref{spectrum}) can be performed analytically, yielding:
\begin{eqnarray}
S(E_\mathrm{vis}) &=& \varepsilon(E_\mathrm{vis})
\int_{E_T}^\infty dE 
\left(\sum_i \;{\cal N}_i \;\Phi_i(E) P_i(E)\right) \frac{\sigma(E)}{E_2-E_1}\;
\int_{E_1}^{E_2} dE_e \;
r(E_e+m_e,\,E_\mathrm{vis})\ \\
&=& \varepsilon(E_\mathrm{vis})
\int_{E_T}^\infty dE 
\left(\sum_i \;{\cal N}_i \;\Phi_i(E) P_i(E)\right)\;\sigma(E) \; R(E,\,E_\mathrm{vis})\ 
\label{spectrum2}
\end{eqnarray}
where $R$ is the recoil-corrected energy resolution function, 
\begin{equation}
R(E,\,E_\mathrm{vis}) =  \frac{1}{2(E_2-E_1)}\left[
\mathrm{erf}\left(\frac{E_2+m_e-E_{\mathrm{vis}}}{\sqrt{2}\sigma_e}\right)-
\mathrm{erf}\left(\frac{E_1+m_e-E_{\mathrm{vis}}}{\sqrt{2}\sigma_e}\right)
\right]\ ,
\label{recoilres}
\end{equation}
with erf$(x)$ defined as \cite{Inte}
\begin{equation}
\mathrm{erf}(x)=\frac{2}{\sqrt{\pi}}\int_0^x dt\, e^{-t^2}\ .
\end{equation}
The function $R$ in Eq.~(\ref{recoilres}) reduces to the function $r$ in Eq.~(\ref{gaussian}) in the recoilless limit.

Figure~3 compares the energy resolution functions with recoil ($R$) and without recoil ($r$), as solid and dotted lines, respectively,
for different neutrino energies $E$. All functions are aligned to their average visible energy, which is also the origin of the $x$-axis scale $\Delta E_\mathrm{vis}$. The alignment removes one of the recoil effects [the relative displacement of centroids at $O(E/m_p)$] in order to emphasize the other effect, namely, the widening of the energy resolution tails. 

Summarizing, nucleon recoil effects can be implemented in the unbinned spectrum $S$ by using the modified energy resolution function $R$ in Eq.~(\ref{recoilres}), instead of the usual function $r$ in Eq.~(\ref{gaussian}). Similar results hold for a binned spectrum as described below.

\subsection{Recoil effects in binned spectra}

Although we shall focus on unbinned spectral analyses in Sec.~V~C, for completeness we also discuss recoil effects in binned spectra, in the realistic case where the efficiency function $\varepsilon(E_\mathrm{vis})$ is smooth enough to be nearly constant in each bin. Let us consider a spectrum $S$ divided into bins, the $i$-th one covering a range $E_\mathrm{vis}\in [E'_i,\,E''_i]$ and containing a number of events given by 
\begin{equation}
N_i =\int_{E'_i}^{E''_i}dE_\mathrm{vis}\; S(E_\mathrm{vis})\ .
\end{equation}
Since $S$ is a double integral [see Eq.~(\ref{spectrum})], the calculation of $N_i$ involves in general a triple integral, $\int dE_\mathrm{vis} \int dE_e \int dE$. A useful reduction is possible if the efficiency function $\varepsilon(E_\mathrm{vis})$ can be taken as approximately constant in each bin range, namely, $\varepsilon(E_\mathrm{vis})\simeq \varepsilon_i$ for
$E_\mathrm{vis}\in [E'_i,\,E''_i]$. In this case, the integration ordering can be swapped into $\int dE \int dE_e \int dE_\mathrm{vis}$, where the two inner integrals are analytical. A similar reduction was used in \cite{Mont} in another context. The final result is:
\begin{equation}
N_i = \varepsilon_i \int_{E_T}^\infty dE 
\left(\sum_i \;{\cal N}_i \;\Phi_i(E) P_i(E)\right)\; \sigma(E) \; W_i(E)\ ,
\label{bin}
\end{equation}
where the function $W_i(E)$ is given by
\begin{equation}
W_i(E) =\frac{\sqrt{2}\sigma_e}{2(E_2-E_1)}\left[
g(E''_i-E_1)-g(E''_i-E_2)-g(E'_i-E_1)+g(E'_i-E_2)\right]\ ,
\end{equation}
and
\begin{equation}
g(x) = 
\frac{x-m_e}{\sqrt{2}\sigma_e}\,\mathrm{erf}\left(\frac{x-m_e}{\sqrt{2}\sigma_e}\right)
+\frac{1}{\sqrt{\pi}}\,e^{-\left(\frac{x-m_e}{\sqrt{2}\sigma_e}\right)^2}\ .
\end{equation}
In the above formulae, tiny variations of $\sigma_e$ for $E_e\in[E_1,\,E_2]$ have been neglected [e.g., the value of $\sigma_e$ can be taken at $E_e=(E_1+E_2)/2$].  
In the limit of no recoil and perfect resolution ($\sigma_e\to 0$),  $W_i$ reduces to a top-hat function of width $E''_i-E'_i$; finite resolution and recoil effects smear out the top-hat shape. 
In conclusion, with or without binning, recoil effects on the event spectrum can be included in terms of a single integral over the neutrino energy $E$ with appropriate kernels, according to Eqs.~(\ref{spectrum2}) and (\ref{bin}).

\section{Oscillation probability}   

In this Section we discuss in detail the reactor neutrino survival probability
$P_{ee}$. We cast $P_{ee}$ in a closed analytical form, including matter and 
multiple reactor effects [see Eq.~(\ref{P3numat}) below]. This form allows to make the
discrete parameter $\alpha$  in Eq.~(\ref{alpha}) continuous, so as to interpolate
smoothly between NH ($\alpha=+1$) and IH ($\alpha=-1$). In this way one can cover the null case of  
``undecidable'' hierarchy ($\alpha=0$) in the subsequent statistical analysis.

\subsection{Oscillation probability in vacuum}

Using the notation in Sec.~II, the $3\nu$ vacuum survival probability $P(\overline\nu_e\to\overline\nu_e)$ can be written in the form 
\begin{equation}
P^{3\nu}_\mathrm{vac}=1-4c^4_{13}s^2_{12}c^2_{12}\sin^2\delta - 4s^2_{13}c^2_{13}c^2_{12}\sin^2(\alpha \Delta + \delta/2)-
4s^2_{13}c^2_{13}s^2_{12}\sin^2(-\alpha \Delta + \delta/2)\ .
\label{P3vac}
\end{equation}
As observed in \cite{Fo01}, the above expression is not
invariant under a change of hierarchy $(\alpha \to -\alpha)$ , except for the case 
$c^2_{12}=s^2_{12}$ which is experimentally excluded.

It is tempting to separate $\alpha$-odd terms in the oscillation {\em amplitudes}. 
However, these terms carry a spurious dependence on the conventional squared mass parameter 
which is kept fixed while its sign is flipped. For instance,
$\alpha$-odd terms at fixed $\Delta m^2$ in Eq.~(\ref{P3vac}) are proportional to $\sin \delta$, 
\begin{equation}
P^{3\nu}_\mathrm{odd} = 2\,\alpha\, s^2_{13}c^2_{13}(s^2_{12}-c^2_{12})\sin (2 \Delta) \sin\delta\ ,
\label{wrong}
\end{equation}
while $\alpha$-odd terms at fixed $\Delta m^2_{31}$ \cite{Hagi} or fixed $\Delta m^2_{32}$ \cite{LoSe}
are proportional to $\sin 2\delta$. 
Convention-independent effects should not impose that the largest squared mass difference 
(be it $\Delta m^2_{31}$, or $\Delta m^2_{32}$, or a combination such as $\Delta m^2$) is the same 
in NH and IH. It is thus incorrect to claim, on this basis, that $\sin 2\delta =1$ is 
an optimal condition to observe hierarchy effects in reactor experiments \cite{LoSe}.

\begin{figure}[t]
\vspace*{.0cm}
\hspace*{.0cm}
\includegraphics[scale=0.48]{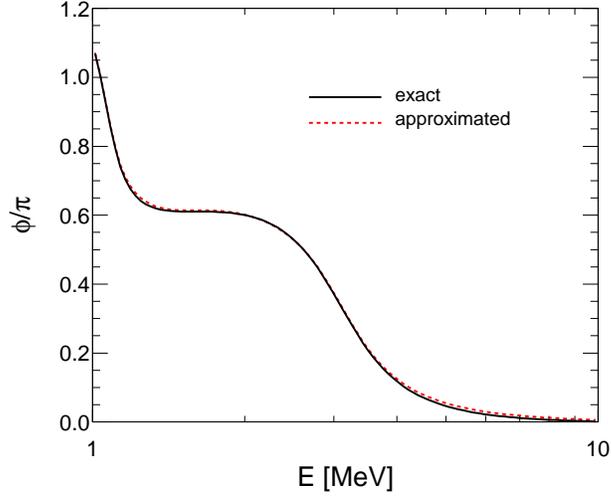}
\vspace*{0.0cm}
\caption{\label{f04}
Comparison of exact and approximate values (in units of $\pi$) of the phase contribution $\varphi$ embedding hierarchy effects, as
a function of neutrino energy $E$, for $s^2_{12}=0.307$, $\delta m^2=7.54\times 10^{-5}$~eV$^2$, and $L=52.5$~km. 
See the text for details.}
\end{figure}

In order to circumvent this drawback, one may separate $\alpha$-odd terms in the oscillation {\em phase\/}  
without fixing the squared mass parameter, as proposed in \cite{Nuno,Mina} and  
revisited in \cite{Ciu0,Voge}. In particular, the probability $P^{3\nu}_\mathrm{vac}$ in Eq.~(\ref{P3vac}) 
can be exactly rewritten as \cite{Mina}:
\begin{equation}
P^{3\nu}_\mathrm{vac}=c^4_{13}P^{2\nu}_\mathrm{vac} + s^4_{13} +
2s^2_{13}c^2_{13}\sqrt{P^{2\nu}_\mathrm{vac}}\cos(2\Delta_{ee}+\alpha \varphi)\ ,
\label{P3nuphase}
\end{equation}
in terms of the $2\nu$ limit 
\begin{equation}
P^{2\nu}_\mathrm{vac}=\lim_{\theta_{13}\to 0} P^{3\nu}_\mathrm{vac} = 1-4s^2_{12}c^2_{12}\sin^2\delta\ ,
\label{P2nu}
\end{equation}
and of an effective squared mass parameter \cite{Gouv,Nuno,Mina}, 
\begin{equation}
\Delta m^2_{ee} = \Delta m^2 +\frac{\alpha}{2}(c^2_{12}-s^2_{12})\delta m^2\ , 
\label{Deltam2ee}
\end{equation}
with
\begin{equation}
\Delta_{ee}=\frac{\Delta m^2_{ee}L}{4E}=
\Delta +\frac{\alpha}{2}(c^2_{12}-s^2_{12})\delta\ ,
\label{Deltaee}
\end{equation}
while 
the phase $\varphi$ in Eq.~(\ref{P3nuphase}) is parametrically defined as \cite{Mina,Mi12} 
\begin{eqnarray}
\cos\varphi &=& \frac{c^2_{12}\cos(2s^2_{12}\delta)+s^2_{12}\cos(2c^2_{12}\delta)}{\sqrt{P^{2\nu}_\mathrm{vac}}}
\label{cosphi}\ ,\\
\sin\varphi &=& \frac{c^2_{12}\sin(2s^2_{12}\delta)-s^2_{12}\sin(2c^2_{12}\delta)}{\sqrt{P^{2\nu}_\mathrm{vac}}}
\label{sinphi}\ .
\end{eqnarray}
Equation~({\ref{P3nuphase}}) also allows a clear separation  between ``fast'' ($\Delta_{ee}$-driven) oscillations
and ``slow'' ($\delta$-driven) modulations in $P^{2\nu}_\mathrm{vac}$ and $\varphi$.

Expressing $\varphi$ via an $\arctan$ function [from the ratio of Eqs.~(\ref{sinphi}) and (\ref{cosphi})] is not
particularly convenient as it leads to a quadrant ambiguity. 
We have found a useful empirical approximation to $\varphi$ in closed form,
\begin{equation}
\varphi \simeq 2s^2_{12}\delta \left( 
1-\frac{\sin \delta}{2\delta \sqrt{P^{2\nu}_\mathrm{vac}}}\right)\ ,
\label{phiapprox}
\end{equation}
which will be used hereafter. Figure~4 shows a comparison of exact and approximate values of $\varphi$ as a function of neutrino energy $E$, calculated for
reference values $s^2_{12}=0.307$, $\delta m^2=7.54\times 10^{-5}$~eV$^2$, and $L=52.5$~km. The numerical differences are 
negligible for any practical purpose. 
Similar results (not shown) hold for $s^2_{12}$ and $\delta m^2$ taken in their
$\pm3\sigma$ phenomenological range \cite{Fo12}.
In addition, the approximate expression for $\varphi$ [Eq.~(\ref{phiapprox})] 
shares two analytical properties of the exact parametric definition of $\varphi$ 
[Eqs.~(\ref{cosphi}) and (\ref{sinphi})], namely:  it
periodically increases with $\delta$ as $\varphi(\delta +\pi)=\varphi(\delta)+2\pi s^2_{12}$ \cite{Mina},
and it starts with a cubic term ($\delta^3$) in a power expansion \cite{Ciu0}.

As it was emphasized in \cite{Mina} and later in \cite{Ciu0,Voge}, 
the hierarchy dependence of $P^{3\nu}_\mathrm{vac}$ is physically manifest in the odd term 
$\pm\varphi$, which induces either an observable advancement ($+\varphi$) or a retardation ($-\varphi$) of the oscillation phase, with
a peculiar energy dependence {\em not\/} proportional to $L/E$ (see Fig.~4). 
Conversely, hierarchy-odd effect which are proportional to $L/E$ [as in Eq.~(\ref{Deltaee})]
are immaterial, as far as they can be absorbed into a redefinition of $\Delta m^2$
within experimental uncertainties. 
 Determining the hierarchy with reactor experiments thus amounts to finding evidence for an extra, non-$L/E$ oscillation
phase with definite sign (either $+\varphi$ or $-\varphi$), for unconstrained values of $\Delta m^2_{ee}$. This requirement 
places the focus of the measurement on the low-energy part of the spectrum where $\varphi$ is large, while
the high-energy part acts as a calibration.

\subsection{Multiple reactor cores}

In the presence of $n=1,\dots,N$ reactor cores (placed at slightly different distances $L_n$ and contributing with 
different fluxes $\Phi_n)$, damping effects arise on the fast oscillating terms, while being negligible on 
the slow ones \cite{Ciu2,DYB2}. Such effects can be taken into account analytically as follows.

Let us define the flux weights $w_n$, the flux-weighted baseline $L$, and the fractional baseline differences $\lambda_n$ as
\begin{eqnarray}
w_n &=& \frac{\Phi_n}{\sum_n \Phi_n}\ ,\\
L   &=& \sum_n w_n\, L_n \ ,\\
\lambda_n &=& \frac{L_n - L}{L}\ ,
\end{eqnarray}
where $\sum_nw_n=1$ and $\sum_n \lambda_n=0$. The fast oscillating term in $P^{3\nu}_\mathrm{vac}$ is
obtained by summing up the weighted contributions from different cores, 
\begin{equation}
P^{3\nu}_\mathrm{vac} \simeq c^4_{13}P^{2\nu}_\mathrm{vac} + s^4_{13} + 2 s^2_{13}c^2_{13}\sqrt{P^{2\nu}_\mathrm{vac}} \sum_n w_n
\cos\left(\frac{\Delta m^2_{ee}L_n}{2E}+\alpha \varphi\right)\ ,
\end{equation}
and by reducing it via the trigonometric identity
\begin{equation}
\sum_n w_n \cos(x+\xi_n)=w\cos(x+\xi)\ ,
\end{equation}
where 
\begin{eqnarray}
w^2 &=& \sum_{n,m}w_n\,w_m\cos(\xi_n-\xi_m)\ ,\\
\tan\xi &=& \frac{\sum_n w_n\sin\xi_n}{\sum_nw_n\cos\xi_n}\ . 
\end{eqnarray}
In our case, $x=(\Delta m^2_{ee}L/2E)+\alpha\varphi$ and $\xi_n=\Delta m^2_{ee}L\lambda_n/2E$. 
By keeping the first nontrivial terms in a $\xi$ and $\xi_n$ power expansion, the final result can be cast
in the form 
\begin{equation}
P^{3\nu}_\mathrm{vac} \simeq c^4_{13}P^{2\nu}_\mathrm{vac} + s^4_{13} + 2 s^2_{13}c^2_{13}\sqrt{P^{2\nu}_\mathrm{vac}} \,w\,
\cos(2\Delta_{ee}+\alpha \varphi)\ ,
\label{damped}
\end{equation}
where the damping factor $w$ reads
\begin{equation}
w \simeq 1-2(\Delta_{ee})^2\sum_nw_n\lambda_n^2\ .
\end{equation}

Let us consider the specific JUNO setting, characterized by $N=10$ reactor cores 
(6 being located at Yangjiang and 4 at Taishan) with average power $P_n$ \cite{DYB2}. Assuming fluxes 
$\Phi_n\propto P_n/L_n^2$, we obtain a flux-weigthed distance $L=52.474$~km and a damping coefficient 
$\sum_n w _n \lambda_n^2 = 2.16\times 10^{-5}$. In this case, the amplitude of the hierarchy-sensitive cosine term in Eq.~(\ref{damped}) 
is reduced by as much as 28\% at low energy ($E\simeq 2$~MeV).

Finally, we remark that damping effects may acquire a slight  
time dependence via reactor power variations, $P_n=P_n(t)$. This dependence may be effectively
embedded in time-dependent weights $w_n=w_n(t)$, baseline $L=L(t)$ and damping factor $w=w(t)$. 
For the sake of simplicity, we shall only consider stationary conditions (constant $L$ and $w$) hereafter.

\subsection{Oscillation probability in matter}

At medium baselines $L\sim O(50)$~km, reactor $\overline\nu_e$ mostly propagate within the upper part of the Earth's crust. For a nearly constant electron density $N_e$, the ratio of matter to vacuum terms in the propagation hamiltonian reads \cite{Blen}
\begin{equation}
\mu_{ij}=\frac{2\sqrt{2}G_F\, N_e\, E}{\Delta m^2_{ij}}=
1.526\times 10^{-7}\left(\frac{N_e}{\mathrm{mol/cm}^3}\right)\left(\frac{E}{\mathrm{MeV}}\right)
\left(\frac{\mathrm{eV}^2}{\Delta m^2_{ij}}\right)\ .
\end{equation}

Assuming a typical crust density $N_e\simeq 1.3$~mol/cm$^3$, the only non-negligible ratio is $\mu_{12}\sim O(10^{-2})$.
Correspondingly, the $(\nu_1,\,\nu_2)$ mass-mixing parameters in matter ($\delta \tilde m^2,\,\tilde\theta_{12}$) \cite{Blen}
read, at first order in $\mu_{12}$ and for $\overline\nu_e$ oscillations,
\begin{eqnarray}
\sin 2\tilde \theta_{12} &\simeq& \sin 2\theta_{12}(1-\mu_{12}\cos2\theta_{12})\ ,\label{p1}\\
\delta \tilde m^2 &\simeq& \delta m^2(1+\mu_{12}\cos2\theta_{12})\ .\label{p2}
\end{eqnarray}
Note that, for $E\sim 8$~MeV, the fractional matter correction to 
mass-mixing parameters is $\sim 8\times 10^{-3}$, which is definitely not negligible 
as compared with the prospective fit accuracy on the same parameters (see below). 

We implement matter effects via the replacement $(\delta m^2,\,\theta_{12})\to(\delta \tilde m^2,\,\tilde\theta_{12})$ from 
Eqs.~(\ref{p1},\ref{p2})
into $P^{2\nu}_\mathrm{vac}$, obtaining as a final result
\begin{equation}
P^{3\nu}_\mathrm{mat} \simeq c^4_{13}P^{2\nu}_\mathrm{mat} + s^4_{13} + 2 s^2_{13}c^2_{13}\sqrt{P^{2\nu}_\mathrm{mat}} \,w\,
\cos(2\Delta_{ee}+\alpha \varphi)\ ,
\label{P3numat}
\end{equation}
where
\begin{equation}
P^{2\nu}_\mathrm{mat}= 1-4\tilde s^2_{12}\tilde c^2_{12}\sin^2\tilde \delta\ .
\label{P2numat}
\end{equation}
These two equations provide our ``master formula'' for the oscillation probability in either NH ($\alpha=+1$)
or IH ($\alpha=-1$), including matter effects in the crust and damping effects of multiple reactor cores.

A final remark is in order. We have omitted the replacement $(\delta m^2,\,\theta_{12})\to(\delta \tilde m^2,\,\tilde\theta_{12})$ into $\varphi$, since it leads to insignificant numerical variations of $P^{3\nu}_\mathrm{mat}$.
We have also compared the above $P^{3\nu}_\mathrm{mat}$ with the exact probability derived from 
numerical flavor evolution in matter of $\overline\nu_e$'s from each single reactor source,
\begin{equation}
P^{3\nu}_\mathrm{exact} = \sum_n w_n\,P^{3\nu}_\mathrm{exact}(L_n,\,E,\,N_e,\,\delta m^2,\,\Delta m^2_{ee},\,
\theta_{12},\,\theta_{13},\,\alpha)\ ,
\label{P3nuexact}
\end{equation}
for $\alpha=\pm1$, obtaining   permill-level differences 
($|P^{3\nu}_\mathrm{mat}-P^{3\nu}_\mathrm{exact}|< 2\times 10^{-3}$ for $E\geq E_T$) which can be safely neglected in the data
analysis. In conclusion, Eq.~(\ref{P3numat}) is a very good approximation to the exact oscillation probability.

\subsection{Continuous ``interpolation'' between the two hierarchies}

The analytical form of $P_{ee}$ in Eq.~(\ref{P3numat}) isolates
hierarchy effects via an extra, non-$L/E$ contribution 
$\pm\varphi$ to the ``fast'' $L/E$ oscillation phase  $2\Delta_{ee}$;
then, the sign of the extra phase (i.e., the occurrence of either an advancement or a retardation of phase) 
can tell the hierarchy \cite{Mina,Ciu0,Voge}.

In realistic situations, it may occur that an experiment finds  
no evidence for an extra phase, or some evidence for it but with the wrong sign,
or even with a wrong (too large or too small) amplitude. We propose to cover all
these possible outcomes by generalizing the discrete parameter $\alpha=\pm 1$ in
Eqs.~(\ref{alpha}) and (\ref{P3numat}) as a formally continuous parameter,  
\begin{equation}
\alpha =\pm 1\ \rightarrow\ \alpha = \mathrm{free \ parameter}\ ,
\end{equation}
whose value should be constrained by a fit to prospective or real data. Evidence for  $\alpha\neq 0$ will
then translate into evidence for hierarchy effects, with NH or IH being signaled by
sign($\alpha$). Viceversa, the hierarchy discrimination will be
compromised, if the data favor either the ``null'' case
$\alpha\simeq 0$, or implausible cases with $|\alpha|> 1 $.

This phenomenological approach makes the data analysis easier, since $\alpha$ is formally treated as any other 
free parameter in the fit; moreover, it 
offers an alternative viewpoint to some subtle statistical 
issues recently highlighted in \cite{Sta1,Hagi,Sta2}.  In particular, it appears that the traditional
$\Delta\chi^2$ distance between the ``true'' and ``wrong'' hierarchy cases, if naively interpreted, may overestimate
the real sensitivity to the hierarchy
for at least two reasons:  (1) $\Delta\chi^2$ is an
appropriate statistical measure for (continuous) parameter estimation tests, but not
necessarily for (discrete) hypotheses tests; (2) the hierarchy discrimination is already 
compromised in cases which are half-way between the ``true'' and ``wrong'' expectations. 
Various statistical methods and measures have been introduced in \cite{Zha1,Zha2,Sta1,Hagi,Sta2}  
to quantify more properly the hierarchy sensitivity.  
Our approach offers an alternative perspective by (1) introducing
a continuous parameter $\alpha$ which allows usual $\chi^2$ analyses, and (2) 
comparing the cases $\alpha=\pm1$ with the null case $\alpha=0$ in order to estimate the hierarchy sensitivity.
See Sec.~VI for related comments.

\subsection{Oscillation probability for geoneutrinos and far reactors}

In general, medium-baseline reactor experiments designed to probe the hierarchy at $L\sim O(50)$~km suffer from
irreducible backgrounds from farther reactors at $L\gg 50$~km \cite{DYB2,Ciu0} (insensitive to $\Delta m^2$) 
and from geoneutrinos \cite{Baty} (insensitive to both $\Delta m^2$ and $\delta m^2$). 
For the ``far'' and ``geo'' background components we shall take the oscillation probability as 
\begin{equation}
P^{3\nu}_\mathrm{far} \simeq c^4_{13}P^{2\nu}_\mathrm{mat}+s^4_{13}\ ,
\end{equation}
with $P^{2\nu}_\mathrm{mat}$ as in Eq.~(\ref{P2numat}),
and
\begin{equation}
P^{3\nu}_\mathrm{geo} \simeq c^4_{13}(1-2s^2_{12}c^2_{12})+s^4_{13}\ ,
\end{equation}
respectively.

We remark that the geoneutrino background may acquire a slight $\delta m^2$ dependence
through non-averaged oscillation effects in the local crust. These effects, not considered herein, may be estimated or 
at least constrained
by constructing detailed geological models for the local distribution of Th and U geoneutrino sources
\cite{Mant}.

\section{Ingredients of the numerical and statistical analysis} 

In the previous Sections~I and II we have discussed features of the differential cross section and of the
oscillation probabilities, which may be useful for generic medium-baseline reactor experiments. 
In this section we describe further ingredients which refer  to the specific 
JUNO experimental setting described in \cite{DYB2} and to other choices made in our numerical 
and statistical analysis.

\subsection{Priors on oscillation parameters}

At present, global neutrino data analyses show no significant indication in favor of either
normal or inverted hierarchy. We thus conflate the (slightly different) current results for normal and inverted
hierarchy as taken from \cite{Fo12}, and assume the following $\alpha$-independent priors for the relevant oscillation parameters 
in Eq.~(\ref{P3numat}),
\begin{eqnarray}
\delta m^2/\mathrm{eV}^2 &=& (7.54\pm0.24) \times 10^{-5}\ ,\label{par1}\\
\Delta m^2_{ee}/ \mathrm{eV}^2 &=& (2.43\pm0.07) \times 10^{-3}\ ,\label{par2}\\
s^2_{12} &=& 0.307\pm 0.017\ ,\label{par3}\\
s^2_{13} &=& 0.0242\pm 0.0025\label{par4}\ ,
\end{eqnarray}
with errors at $\pm1\sigma$.

\subsection{Fluxes and normalization}

In this section we fix the fluxes and normalization factors in the integrand of Eq.~(\ref{spectrum}), namely,
\begin{equation}
{\cal N}_\mathrm{MB}\Phi_\mathrm{MB}P^{3\nu}_\mathrm{mat}
+{\cal N}_\mathrm{far}\Phi_\mathrm{far}P^{3\nu}_\mathrm{far}
+{\cal N}_\mathrm{geo}\Phi_\mathrm{geo}P^{3\nu}_\mathrm{geo}
\label{normalize}
\end{equation}
where, in the context of JUNO, the three terms refer to the contributions from 
the 10 medium-baseline reactors (MB) \cite{DYB2}, the two dominant far-reactor complexes (far) \cite{DYB2},
and geoneutrinos (geo) \cite{Geon}, respectively.

The reactor fluxes depend, in general, on the (time-dependent) relative U and Pu fuel components. 
For our prospective data analysis, we assume typical average values from Fig.~21 in \cite{DB02}, 
\begin{equation}
{}^{235}\mathrm{U}:{}^{239}\mathrm{Pu}:{}^{238}\mathrm{U}:{}^{241}\mathrm{Pu}\,
 \simeq\, 0.60:0.27:0.07:0.06\ ,
\end{equation}
for both medium-baseline and far reactors. The corresponding fluxes are taken from \cite{Hube}. 

Concerning the reactor event normalization, from the information reported in \cite{DB02}
 we derive the following rough estimate for the number 
of unoscillated events, expected for a detector of mass $M$ at distance $L$ from a reactor complex of thermal power $P$ in typical
conditions at Daya Bay (including detection efficiencies and reactor duty cycles): 
\begin{equation}
\frac{\mathrm{unoscillated\ events}}{\mathrm{year}}\simeq 2.65 \times 10^5
\left(\frac{M}{\mathrm{kT}}\right)
\left(\frac{P}{\mathrm{GW}}\right)
\left(\frac{\mathrm{km}}{L}\right)^2\ .
\end{equation}
For our numerical analysis of JUNO, we assume $M=20$~kT and $P=35.8$~GW from \cite{DYB2}, $L=52.474$~km from Sec~IV~D, and 
an exposure of 5 years, 
yielding a total of $3.4\times 10^{5}$ events expected for no oscillations; these numbers fix the normalization of the term  
${\cal N}_\mathrm{MB}\Phi_\mathrm{MB}$ after energy integration. Oscillations typically reduce the expectations to $\sim 10^{5}$ events
for oscillation parameters as in Sec.~V~A,
hence the title of this work. Such an oscillated rate corresponds to $\sim 55$ oscillated events per day in typical
conditions.%
\footnote{Our estimate seems more optimistic than the rate of $\sim 40$ events/day quoted in \protect\cite{SanF}. 
We are unable to trace the source(s) of this difference which, if confirmed, could be compensated by rescaling our assumed 
lifetime from 5 to 6.8 years in order to collect the same event statistics.} 

By repeating the previous exercise for the two far reactors with 
power $P=17.4$~GW at $L=215$~km and 265~km \cite{DYB2}, we obtain 
$10^4$ and $6.5\times 10^3$ unoscillated events in five years, respectively. These estimates fix the normalizations of 
the two far-reactor subterms in ${\cal N}_\mathrm{far}\Phi_\mathrm{far}$.

\begin{figure}[t]
\vspace*{.0cm}
\hspace*{.0cm}
\includegraphics[scale=0.45]{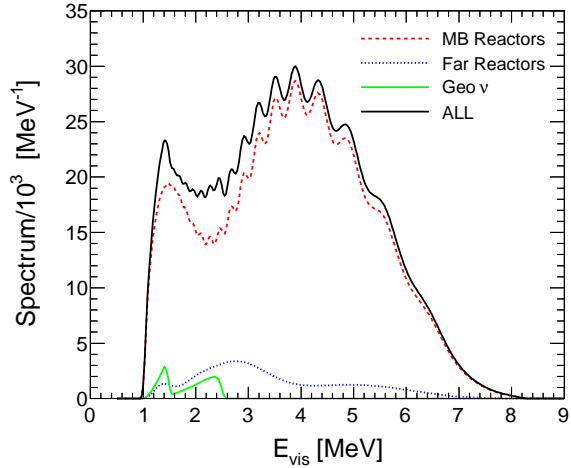}
\vspace*{0.0cm}
\caption{\label{f05}
Absolute energy spectrum of 
events expected in JUNO for normal hierarchy ($\alpha=+1$) and assuming the central values of the oscillation parameters defined in the
text. The breakdown of the total spectrum in its three components (medium baseline reactors, far reactors, geoneutrinos) is also shown}
\end{figure}

Concerning the normalization of geoneutrino events, we assume from \cite{Geon} 
the following unoscillated flux estimates near the Daya Bay site
(central values): $\Phi(\mathrm{U})=4.04\times 10^6$/cm$^2$/s and 
 $\Phi(\mathrm{Th})=3.72\times 10^6$/cm$^2$/s, which correspond to unoscillated event rates 
$R(\mathrm{U})=51.7$~TNU and  $R(\mathrm{Th})=15.0$~TNU, where one terrestrial neutrino unit (TNU) corresponds
to $10^{-32}$ events per target proton per year \cite{Liss}. Assuming a liquid scintillator detector of 20~kT mass and
proton fraction $\sim 11\%$, operating for five years with typical low-energy efficiency $\varepsilon\simeq 0.8$, we estimate an effective geo-neutrino exposure of $\sim 5.2\times 10^{33}$~in units of protons$\times$years, which implies 
$\sim 2.7\times 10^3$ (U) and $\sim 0.8\times 10^3$ (Th) unoscillated events, fixing the geoneutrino
normalization in our analysis. Concerning the geoneutrino fluxes, we use the same spectral shape as in \cite{Mant}. 

Notice that, in the above estimates, typical efficiency factors are already embedded in the normalization factors $\cal{N}$. 
Therefore, we  
take $\varepsilon(E_\mathrm{vis})=1$ in Eq.~(\ref{spectrum}). With all the ingredients described so far, 
the absolute event spectrum can be calculated
for any value of the continuous parameters ($\delta m^2,\,\Delta m^2_{ee},\,\theta_{12},\,\theta_{13},\,\alpha$).%
\footnote{In this study we have ignored further oscillation-independent backgrounds, see \protect\cite{Snow} for a recent
evaluation in the context of JUNO.}

Figure~5 shows the total absolute spectrum of oscillated events and its breakdown into 
three main components (medium-baseline reactors,
far reactors, and geoneutrinos), in terms of the measured visible energy $E_\mathrm{vis}$. The calculation
refers to normal hierarchy ($\alpha=+1$) and to the central values in Sec.~V~A.
Although the far-reactor component is small, its modulation over the whole energy spectrum affects the determination of the 
($\delta m^2,\,\theta_{12}$) parameters which govern the ``slow'' oscillations. In addition, the small 
geoneutrino component adds some ``noise'' at low energy, where most of the hierarchy information is 
confined via the phase $\varphi$. These effects will be discussed quantitatively in Sec.~VI.

For the sake of completeness, Fig.~6 compares the total absolute spectra of oscillated events in the two cases of normal hierarchy ($\alpha=+1$) and inverted hierarchy ($\alpha=-1$). In this figure we have used the same oscillation parameters as in Fig.~5 for both hierarchies, hence the NH and IH spectra merge at high energy where $\varphi\to 0$. The low-energy differences between the two spectra are generally very small, 
and may become even smaller with 
floating mass-mixing parameters, making a detailed statistical analysis mandatory.

\begin{figure}[b]
\vspace*{-.7cm}
\hspace*{.0cm}
\includegraphics[scale=0.45]{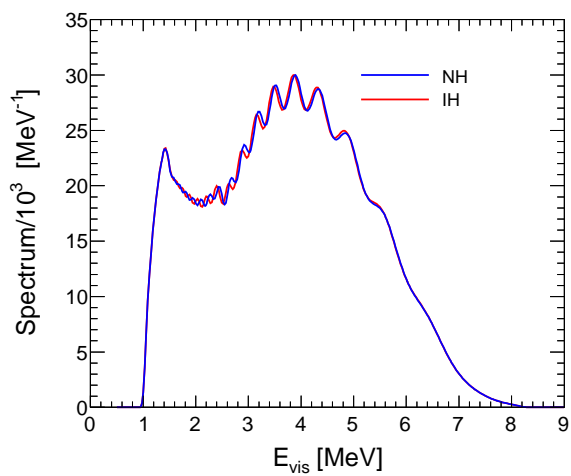}
\vspace*{0.0cm}
\caption{\label{f06}
Comparison of absolute energy spectra of 
events expected in JUNO for normal hierarchy ($\alpha=+1$) and inverted hierarchy ($\alpha=-1$), assuming in both cases the same 
oscillation parameters as in Fig.~5. }
\end{figure}

\subsection{$\chi^2$ function}

We assume that the ``true'' spectrum $S^*(E_\mathrm{vis})$ is the one calculated for the central values
of the oscillation parameters in Sec.~V~A and for either normal hierarchy ($\alpha=+1$) or 
inverted hierarchy ($\alpha=-1$). The ``true'' spectrum $S^*$ is then compared with a family
of spectra $S(E_\mathrm{vis})$
obtained by varying the continuous parameters ($\delta m^2,\,\Delta m^2_{ee},\,\theta_{12},\,\theta_{13},\,\alpha$),
in terms of a $\chi^2$ function which contains statistical, parametric, and systematic components,
\begin{equation}
\chi^2 = \chi^2_\mathrm{stat} + \chi^2_\mathrm{par} + \chi^2_\mathrm{sys}\ .
\end{equation}

Following \cite{Hagi}, we define the statistical component $\chi^2_\mathrm{stat}$ in the limit of ``infinite bins,'' 
\begin{equation}
\chi^2_\mathrm{stat} = \int_{0~\mathrm{MeV}}^{9~\mathrm{MeV}}dE_\mathrm{vis}\,\frac{d\chi^2_\mathrm{stat}}{dE_\mathrm{vis}}=
\int_{0~\mathrm{MeV}}^{9~\mathrm{MeV}}dE_\mathrm{vis}
\left(\frac{S^*(E_\mathrm{vis})-S(E_\mathrm{vis})}{\sqrt{S^*(E_\mathrm{vis})}}\right)^2\ ,
\label{chistat}
\end{equation}
We have verified that this limit is already realized numerically by using $\gtrsim 250$ energy bins, irrespective
of linear or logarithmic binning in $E_\mathrm{vis}$. 

The parametric component $\chi^2_\mathrm{par}$ is a quadratic penalty for the priors on the four oscillation parameters $p_i=\overline p_i\pm\sigma_i$
in Sec.~V~A,
\begin{equation}
\chi^2_\mathrm{par} = \sum_{i=1}^4 \left(\frac{p_i-\overline p_i}{\sigma_i}\right)^2\ .
\label{chipar}
\end{equation}
The continuous parameter $\alpha$, which interpolates between normal hierarchy ($\alpha=+1$)
and inverted hierarchy ($\alpha=-1$) is left free in the fit.

Finally, we assume three systematic normalization  factors $f_j=1$ with $1\sigma$ errors $\pm s_j$ ($j=\mathrm{R},\,\mathrm{U},\,\mathrm{Th}$). The factor $f_R$ multiplies all  (medium-baseline and far) reactor spectra with an assumed error $s_\mathrm{R}=0.03$.
The factors $f_\mathrm{u}$ and $f_\mathrm{Th}$ multiply the U and Th geoneutrino spectra, respectively, with tentative errors 
$s_\mathrm{Th}=s_\mathrm{U}=0.2$. The systematic $\chi^2$ component is then 
\begin{equation}
\chi^2_\mathrm{sys} = \sum_{j=\mathrm{R,U,Th}} \left(\frac{f_j-1}{s_j}\right)^2\ .
\label{chisys}
\end{equation}

In principle, one might include further relevant systematics via appropriate penalties (the ``pull method'' \cite{Pull}).
For instance, energy scale uncertainties and pulls have been introduced in terms of linear  
\cite{Ghos} or even polynomial \cite{DYB2} parameterizations. However, it is not obvious that these parameterizations 
can cover peculiar nonlinear profiles for the energy scale errors \cite{Tesi}, which may mimic the effects  
of the ``wrong hierarchy'' in the worst cases \cite{Voge}. In this context, the issue of systematic shape uncertainties 
is not really captured just by increasing the systematic ``pulls,'' but requires dedicated studies; 
very recent examples have been worked out in \protect\cite{Snow,Ciuf}.
In this work
we prefer to keep $\chi^2_\mathrm{sys}$ as simple as in Eq.~(\ref{chisys}) and to separately discuss the subtle
interplay of energy scale and spectrum shape uncertainties in Sec.~VII.

The total $\chi^2$ used hereafter is  a function of eight parameters, including the $f_j$'s,
\begin{equation}
\chi^2=\chi^2(\delta m^2,\,\Delta m^2_{ee},\,\theta_{12},\,\theta_{13},\,\alpha,\,f_\mathrm{R},\,f_\mathrm{U},\,f_\mathrm{Th})\ .
\end{equation}
Numerically,
the minimization procedure and the identification of isolines of $\Delta\chi^2=\chi^2-\chi^2_\mathrm{min}$ 
is performed through a Markov Chain MonteCarlo method \cite{MCMC}.  By construction, minimization  
yields $\chi^2_\mathrm{min}=0$ when the spectrum $S$ equals the ``true'' one $S^*$. We shall typically show iso-$N_\sigma$ contours, where 
$N_\sigma =\sqrt{\Delta\chi^2}$. Projections of such contours over 
a single parameter provide the bounds at $N_\sigma$  standard deviations on such parameter \cite{PDGR}.
It is understood that undisplayed parameters are marginalized away. 

A final comment is in order. We surmise that, when real data will be available, 
the most powerful statistical analysis will involve maximization of unbinned (or finely binned) 
likelihood in both energy and time domain, as already performed 
in the context of KamLAND results \cite{Time,Koff}. Such an analysis allows to include 
any kind of systematic errors via pulls, and helps to separate, on a statistical basis,
stationary backgrounds (e.g., geoneutrinos) from time-evolving  reactor fluxes, thus enhancing the
statistical significance of the relevant signals \cite{Time,Koff}.
However, a refined time-energy analysis will probably be restricted only to the experimental 
collaboration owning the data, since the detailed reactor core evolution information is generally either 
classified or averaged over long (yearly or monthly) time periods.

\section{Results of the analysis}

We discuss below the results of our statistical analysis of prospective JUNO data as defined in the previous
Section. We focus on the case of true NH, the results for true IH being rather symmetrical.

\begin{figure}[t]
\vspace*{.0cm}
\hspace*{.0cm}
\includegraphics[scale=0.35]{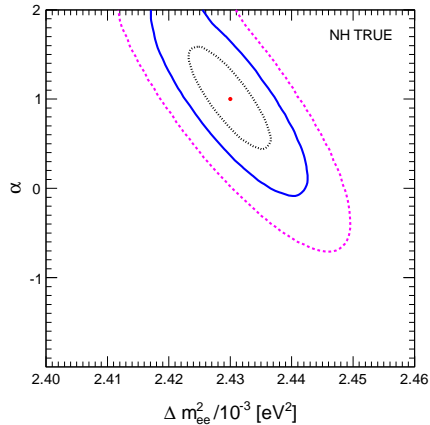}
\vspace*{0.0cm}
\caption{\label{f07} Constraints in the plane $(\Delta m^2_{ee},\,\alpha)$ at 1, 2 and 3$\sigma$ ($\Delta\chi^2=1,\,4,\,9$) from a fit
to prospective JUNO data assuming true normal hierarchy ($\alpha=+1$). Although the inverted hierarchy case ($\alpha=-1$) is $\sim 3.4\sigma$ away, the hierarchy discrimination is already compromised at $\sim 1.7\sigma$, 
where the ``undecidable'' case ($\alpha=0$) is allowed.}
\end{figure}

Figure~7 shows the results of the fit in the plane $(\Delta m^2_{ee},\,\alpha)$ for true NH, in terms
of $N_\sigma= 1,\,2,\,3$ contours for one parameter ($\Delta\chi^2=1,\,4,\,9$), all other parameters
being marginalized away. The errors are rather linear on both parameters, and appear to be significantly
anti-correlated. The anti-correlation stems from the tendency of the fit to keep constant the oscillation
phase $2\Delta_{ee}+\alpha\varphi$ in Eq.~(\ref{P3numat}) for typical neutrino energies $E\simeq 3$--5~MeV:
an increase of $\Delta m^2_{ee}$ is then compensated by a decrease in $\alpha$. 

In Fig.~7, the case of wrong hierarchy ($\alpha=-1$) is formally reached at $\sim 3.4\sigma$; however, it would be misleading 
to take this ``distance'' as a measure of the sensitivity to the hierarchy. Physically, the discrimination of NH vs IH is
successful if the data allow to tell the sign of a non-$L/E$ phase, which advances or retards  
a hierarchy-independent $L/E$ oscillation phase.  In our adopted formalism, this 
requirement amounts to tell the sign of $\alpha$: when the sign is undecidable ($\alpha\simeq 0$), 
the discrimination is already compromised. Therefore, the sensitivity to the hierarchy is more properly
measured by the ``distance'' of the true case (either $\alpha=+1$ or $\alpha=-1$) from the  null case
($\alpha=0$): $N_\sigma\simeq \sqrt{\chi^2(\alpha=\pm 1)-\chi^2(\alpha=0)}$. 
 In Fig.~7, this distance is $\sim 1.7\sigma$, namely, about 1/2 of the $\sim 3.4$ sigma 
which formally separate the NH and IH cases. Thus, we recover independently the approximate 
``factor of two'' reduction of the sensitivity with respect to naive expectations \cite{Qian,Hagi,Ciu4},
as expressed by the ``rule of thumb''  $N_\sigma\simeq 0.5 \sqrt{\Delta\chi^2(\mathrm{NH}-\mathrm{IH})}$ \cite{Snow}.
Our approach reaches such result via a fit to a continuous parameter ($\alpha$), 
which is conveniently treated as any other floating parameters in the statistical analysis. 
The case of true IH (not shown) is very similar, with only a slight enhancement  of the sensitivity  
to the hierarchy ($\sim 1.8\sigma$ instead of $\sim 1.7\sigma$). 

In conclusion, the results in Fig~7 show that the sign of $\alpha$ (i.e., the advancement or retardation
of phase due to the hierarchy) can be determined at a level slightly below $\sim 2\sigma$. This value is
in the same ballpark of all recent estimates under similar assumptions, but has been derived via a different approach.
In particular, we have recovered  the 
``rule of thumb''  $N_\sigma\simeq 0.5 \sqrt{\Delta\chi^2(\mathrm{NH}-\mathrm{IH})}$ that was found and discussed in 
\cite{Qian,Hagi,Ciu4,Snow} for two alternative discrete cases, by connecting the two cases via
a continuous variable $\alpha$, whose sign tells the hierarchy. The hierarchy discrimination is successful 
if the data prefer $|\alpha|=1$ with sufficient significance with respect to $\alpha= 0$; conversely, a preference
for $\alpha=0$ would compromise the experiment, while surprisingly large values $|\alpha|\gg 1$ would signal 
possible systematics which 
are artificially enhancing the hierarchy effects.  If the hierarchy discrimination is successfull, then fit results
such as those in Fig.~7 provide the central value and error of $\Delta m^2_{ee}$ and also of $\Delta m^2$ via Eq.~(\ref{Deltam2ee}), namely:
\begin{equation}
\Delta m^2 = \Delta m^2_{ee} - \frac{\mathrm{sign}(\alpha)}{2}(c^2_{12}-s^2_{12})\delta m^2 \ .
\end{equation}
The determination of the fundamental parameter $\Delta m^2$ thus depend also on the constraints achievable on the 
parameters $(\delta m^2,\,s^2_{12})$, which we now discuss.

\begin{figure}[t]
\vspace*{.0cm}
\hspace*{.0cm}
\includegraphics[scale=0.35]{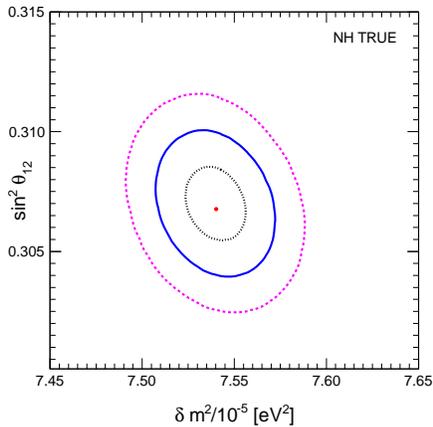}
\vspace*{0.0cm}
\caption{\label{f08} Constraints in the plane $(\delta m^2,\,s^2_{12})$ at 1, 2 and 3$\sigma$ from a fit
to prospective JUNO data, assuming true normal hierarchy.}
\end{figure}

Figure~8 shows the fit results in the plane $(\delta m^2,\,s^2_{12})$ at 1, 2 and 3$\sigma$.  The slight anti-correlation
is due to to the fact that, in general, a slight increase of the ``slow'' oscillation phase $\delta$ can be partly
compensated by a decrease of the corresponding amplitude $\sin^2 (2\theta_{12})$.
The $1\sigma$ errors
on each parameter correspond to a nominal accuracy of a few permill, i.e., one order of magnitude better than
the current experimental constraints. Such a prospective accuracy makes it evident, a posteriori, 
the importance of including sub-percent effects due to propagation in matter (which affect 
both $s^2_{12}$ and $\delta m^2$, see Sec.~IV~C) and to nucleon recoil (which affect $\delta m^2$ via the
$\delta m^2/E$ dependence and the positron energy reconstruction, see Sec.~III).  Energy scale nonlinearities (see next Section)
must also be kept under control at a similar level of accuracy, in order to avoid 
biased determinations of $\delta m^2$. 

Figure~9 shows the prospective constraints in the plane $(f_R,\,s^2_{13})$. The two parameters are positively correlated,
since an increase in the reactor flux normalization can be partly compensated by a higher 
$s^2_{13}$ enhancing electron flavor disappearance. In any case, the improvement on the $s^2_{13}$ accuracy
is moderate: the prior $\sim 10\%$ error on this parameter becomes just $\sim 7\%$ after the fit.

\begin{figure}[b]
\vspace*{.0cm}
\hspace*{.0cm}
\includegraphics[scale=0.35]{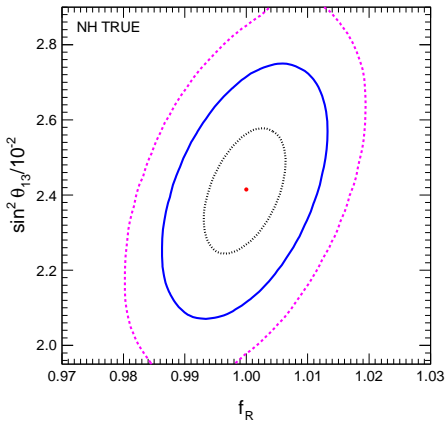}
\vspace*{0.0cm}
\caption{\label{f09} Constraints in the plane $(f_R,\,s^2_{13})$ at 1, 2 and 3$\sigma$ from a fit
to prospective JUNO data assuming true normal hierarchy.}
\end{figure}

Table~I summarizes the information about the parameter accuracy in terms of fractional percent errors at $1\sigma$, before and after the fit to prospective JUNO data, assuming either normal or inverted true hierarchy. In order to
average out small nonlinearities and asymmetries, posterior fractional errors are defined as 1/6 of the $\pm 3\sigma$ fit range, 
divided by the central value of the parameter (which, by construction, is the same before and after the fit).  We
also report the results without the far reactor or geoneutrino backgrounds, so as to gauge their impact on the final accuracy.

Table~I shows that the cases of true NH and IH are almost equivalent in term of final
accuracy on the fit parameters. In particular, $\alpha$ is determined to be $+1.00\pm 0.59$ for true
NH and $-1\pm 0.56$ for true IH. Concerning the other parameters, prospective JUNO data always
lead to a reduction of the prior uncertainty, which is very significant for 
$(\Delta m^2_{ee},\,\delta m^2,\,s^2_{12},\,f_R)$ and moderate for the 
$(s^2_{13},\,f_\mathrm{Th},\,f_\mathrm{U})$. The far-reactor background does not appear to
affect significantly  any fit parameter, while the geoneutrino background and its 
uncertainties  
tend to degrade somewhat the final accuracy of the mass-mixing parameters ($\delta m^2,\,s^2_{12}$),
whose observable oscillation cycle mainly falls in the geoneutrino energy region (see Fig.~5).
Indeed, the ($\delta m^2,\,s^2_{12}$) parameters have non negligible correlations with the
geoneutrino normalization factors $(f_\mathrm{Th},\,f_\mathrm{U})$ after the fit (not shown).

\begin{table}[t]
\caption{\label{Results} Statistical analysis of prospective JUNO data: fractional percent errors ($1\sigma$) 
on the free parameters, before and after the fit to prospective JUNO data, assuming either normal
or inverted true hierarchy. The hypothetical cases without contributions from far reactors (``all $-$ far'')
or from geoneutrinos (``all $-$ geo'') are also reported. In the latter case, the
normalization factors $f_\mathrm{Th,U}$ are absent.}
\begin{ruledtabular}
\begin{tabular}{cccccccc}
Parameter & \% error         	& \multicolumn{3}{c}{\% error after fit (NH true)} 
							& \multicolumn{3}{c}{\% after fit (IH true)}\\
          &  (prior)         & all data & all $-$ far & all $-$ geo & all data & all $-$ far & all $-$ geo \\
\hline
$\alpha$				& $\infty$  	& 59.2	& 59.0 & 57.0   	& 56.2 & 55.3 & 54.0 \\
$\Delta m^2_{ee}$	& 2.0 		& 0.26 	& 0.25 & 0.26 	& 0.26 & 0.25 & 0.25 \\
$\delta m^2$    		& 3.2 		& 0.22 	& 0.21 & 0.16 	& 0.21 & 0.21 & 0.16 \\
$s^2_{12}$			& 5.5 		& 0.49 	& 0.47 & 0.39 	& 0.49 & 0.46 & 0.42 \\
$s^2_{13}$			& 10.3 		& 6.95 	& 6.88 & 6.95 	& 6.84 & 6.77 & 6.84 \\
$f_R$				& 3.0 		& 0.66 	& 0.66 & 0.64 	& 0.65 & 0.65 & 0.64 \\
$f_\mathrm{Th}$		& 20.0 		& 15.3 	& 14.6 & ---  	& 15.5 & 15.4 & --- \\
$f_\mathrm{U}$		& 20.0 		& 13.3 	& 13.3 & ---  	& 13.3 & 13.3 & --- 
\end{tabular}
\end{ruledtabular}
\end{table}

Finally, we discuss the contributions to the $\chi^2$ difference between ``true'' and 
``wrong'' hierarchy, assuming for definiteness the case of true NH as in Fig.~7. The best fit for fixed 
$\alpha=-1$ (wrong hierarchy) is reached at $\chi^2=11.7$, and is dominated by the statistical
contribution ($\chi^2_\mathrm{stat}=11.5$). Figure~10 shows the corresponding $\chi^2_\mathrm{stat}$ density, namely,
the integrand of Eq.~(\ref{chistat}), as function of the visible energy $E_\mathrm{vis}$, together with
its cumulative distribution (i.e., the integral of the density with running upper limit). 
It can be seen that $80\%$ of the contribution to the $\chi^2$  comes from the spectral fit in a very small range
at low energy, $E_\mathrm{vis}\in [1.5,\,3.5]$~MeV. In this range, the vertical mismatch between the true
and wrong spectra changes sign many times, leading to a wavy pattern of the $\chi^2$ density,
also visible with smaller amplitude at higher energies. Intuitively, one can recognize 
that this wavy pattern is very fragile under small relative changes of the horizontal scale
between the true and wrong spectra, due to possible energy scale uncertainties which, in the worst 
cases, might largely erase the pattern itself, at least at low energy. The next Section is
devoted to a discussion of this issue, whose relevance was pointed out in \cite{Voge}.   
 
\begin{figure}[b]
\vspace*{-.2cm}
\hspace*{.0cm}
\includegraphics[scale=0.38]{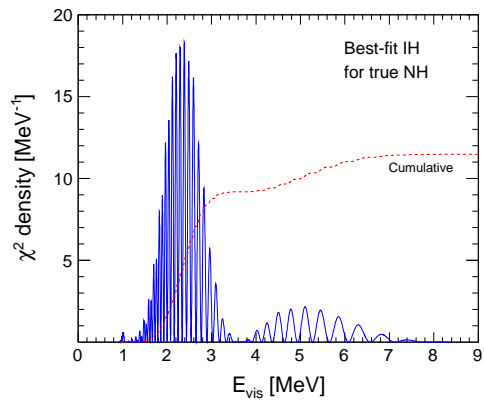}
\vspace*{0.0cm}
\caption{\label{f10} Density and cumulative distribution functions for $\chi^2_\mathrm{stat}$ in the
case of ``wrong'' inverted hierarchy, assuming ``true'' normal hierarchy. The cumulative
function values can be read on the same vertical axis as for the density, but in dimensionless units.}
\end{figure}

\section{Energy scale and spectral shape uncertainties}
 
It was observed in \cite{Voge} that changes in energy scale ($E\to E'$) at percent level can flip the 
sign of the hierarchy-dependent phase $\varphi$ in Eq.~(\ref{P3numat}) (namely, $\alpha=\pm 1\to \alpha=\mp 1$),
provided that 
\begin{equation}
\frac{\Delta m^2_{ee}\,L}{2E}\pm \varphi(E) = \frac{\Delta m^{2\,\prime}_{ee} \,L}{2E'}\mp \varphi(E')
\ ,
\label{Escale}
\end{equation}
where $\Delta m^2_{ee}\neq \Delta m^{2\,\prime}_{ee}$ in general. Multiplying the above terms by 
$2 E'/L\Delta m^2_{ee}$ one gets an implicit equation for the ratio $E'/E$, which is amenable
to iterative solutions by taking 
$\varphi(E)$ as in Eq.~(\ref{phiapprox}). The first iteration after the
trivial solution ($E\simeq E'$) yields the compact expression:
\begin{equation}
\frac{E'}{E}\simeq \frac{\Delta m^{2\prime}_{ee}}{\Delta m^2_{ee}} \mp 2s^2_{12}\frac{\delta m^2}{\Delta m^2_{ee}}
\left(1-\frac{\sin\delta(E)}{2\delta(E)\sqrt{P^{2\nu}_\mathrm{vac}(E)}}\right)\ ,
\label{iter}
\end{equation}
which is already a very good approximation to the exact numerical solution of Eq.~(\ref{Escale}), as we have verified 
in a number of cases. Note that the upper (lower) sign refer to true NH (IH). 
Equation~(\ref{iter}) usefully separates the linear and nonlinear terms 
which can jointly flip the sign of the phase $\varphi$ in the transformation $E\to E'$.

It has been shown that transformations $E\to E'$ as in Eq.~(\ref{iter}) can compromise the hierarchy determination 
\cite{Voge,Snow,Ciuf}, even if they do not lead to a complete degeneracy
between the observable spectra in NH and IH. Below we discuss in detail two specific cases in the 
context of the JUNO project, and then we make some general comments. 

\subsection{Energy scale transformation $E\to E'$ with $E=E'$ at high energy}

Let us specialize Eq.~(\ref{iter}) by selecting the sub-case with 
 $\Delta m^2_{ee}= \Delta m^{2\,\prime}_{ee}$. In this case, the function $E'/E$ takes  the nonlinear form
reported in Fig.~11, where the curves for true NH and IH tend to unit value at high energy
(see also \cite{Voge}). For definiteness, we consider the case of true NH,  the case of true IH being qualitatively very similar. 
Several consequences emerge in the fit, which deserves a detailed discussion.

\begin{figure}[t]
\vspace*{.0cm}
\hspace*{.0cm}
\includegraphics[scale=0.52]{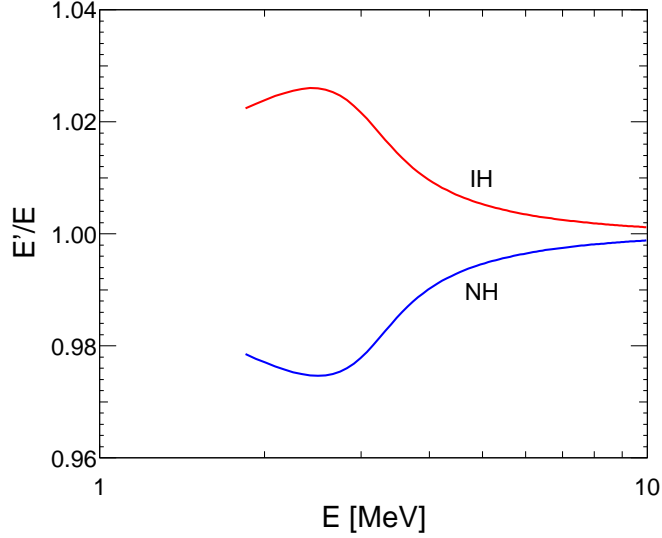}
\vspace*{-0.cm}
\caption{\label{f11} Profile of the neutrino energy ratio $E'/E$ which flips the sign of the
hierarchy-dependent phase $\varphi$ in the JUNO experiment, for the case  $\Delta m^2_{ee}= \Delta m^{2\,\prime}_{ee}$.
The profiles for true NH and true IH are shown for $E\geq E_T=1.806$~MeV.}
\end{figure}

First of all, the parameter $\alpha$ is shifted from the true value $\alpha=+1$ to a wrong
fitted value $\alpha\simeq -1$, as expected from the sign flip of $\varphi$. Figure~12 shows this
shift in the plane $(\Delta m^2_{ee},\,\alpha)$, to be compared with the results in Fig.~7.
It appears the error ellipses are moved downwards from $\alpha=+1$ to $\alpha\simeq -1$
at nearly the same value of $\Delta m^2_{ee}$. 

\begin{figure}[t]
\vspace*{-0.6cm}
\hspace*{.0cm}
\includegraphics[scale=0.39]{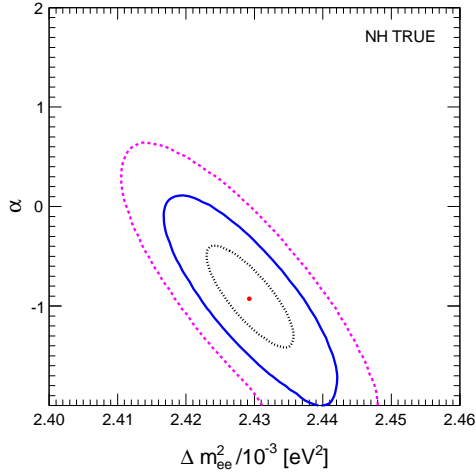}
\vspace*{-0.3cm}
\caption{\label{f12} Constraints in the plane $(\Delta m^2_{ee},\,\alpha)$ for true NH, with 
energy scale variations a in Fig.~11. (Compare with Fig.~7.)}
\end{figure}

However, at the preferred point $(\Delta m^2_{ee},\,\alpha)\simeq (2.43\times 10^{-3}~\mathrm{eV}^2,\,-1)$
in Fig.~12, the best fit is very poor, being characterized by $\chi^2\simeq 360$. The reason is
that, as also recently observed in \cite{Ciuf,Snow}, the degeneracy induced by the transformation
$E\to E'$ is never exact, since it also changes other spectral ingredients besides the
oscillation phase $\varphi$.
In particular, in our numerical experiment, it leads to a noticeable energy shift 
of $\simeq 2.2\%$ close to the energy threshold, as one can read directly from Fig.~11. 
As a result, the rapidly rising 
part of the spectrum just above threshold moves by the same amount, and the 
agreement between expected and observed spectra at low energy is compromised, as it can be seen in  Fig.~13.
The analysis of the $\chi^2$ density in Fig~14 confirms that the energy scale shift $E\to E'$ 
does erase the wavy pattern in Fig.~10 (as a consequence of the sign flip of $\varphi$), 
but it also leads to a large increase of the $\chi^2$ just around the threshold and, to a much lesser 
extent, around the two step-like features of the geoneutrino energy spectrum. Therefore,
the low-energy part of the observed spectrum  may act as a self-calibrating tool to diagnose 
energy scale shifts at percent level near the known IBD threshold ($E_T=1.806$~MeV).%
\footnote{Additional effects (not shown) induced by the energy scale transformation in Fig.~11 include
shifts of best-fit parameters $(\delta m^2,\,s^2_{12})$ and
$(f_\mathrm{U},\,f_\mathrm{Th})$ by $\sim 1\sigma$--$2\sigma$, in units of standard deviations after the fit (see Table~I). Therefore,
energy scale errors tend also to significantly bias such parameters, while the corresponding biases on the 
$(f_R,\,s^2_{13})$ parameters are found to be $<1\sigma$ in our analysis.}

\begin{figure}[b]
\vspace*{0.0cm}
\hspace*{.0cm}
\includegraphics[scale=0.40]{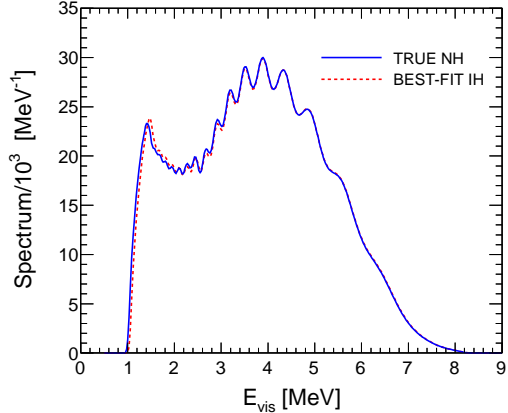}
\vspace*{-0.0cm}
\caption{\label{f13} Comparison of the true NH spectrum with the best-fit IH spectrum from Fig.~12.}
\end{figure}

\begin{figure}[t]
\vspace*{0.0cm}
\hspace*{.0cm}
\includegraphics[scale=0.36]{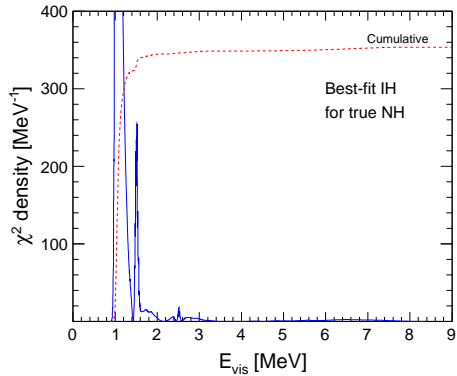}
\vspace*{-0.0cm}
\caption{\label{f14} $\chi^2$ density and its cumulative distribution for the
best-fit IH point in Fig.~12. At low energy, the density function reaches very high values which are partly out of scale.}
\end{figure}

However, the possible self-calibration of the low-energy spectrum tail may fail, if 
the spectral shape  itself is not accurately known in that region. In particular, if the shape errors
in the observable reactor spectrum $\Phi\sigma$ (where $\Phi$ is the reactor flux and $\sigma$ is the IBD cross section)
are comparable to the deviations $\Phi(E)\sigma(E)\to \Phi(E')\sigma(E')$
induced by $E\to E'$, 
then most of the low-energy spectral changes can be ``undone'' by a fudge factor $f(E)$ with the following energy profile: 
\begin{equation}
f(E) = \frac{\Phi(E)\sigma(E)}{\Phi(E')\sigma(E')}\ ,
\label{fudge}
\end{equation}
in which case the best-fit spectrum for the wrong hierarchy becomes 
almost completely indistinguishable from the original, true-hierarchy spectrum. 
In other words, the simultaneous occurrence of an energy scale transformation as in 
Eq.~(\ref{iter}) and of a spectral deformation as in Eq.~(\ref{fudge}) make the true and wrong
hierarchies nearly degenerate from a phenomenological viewpoint. 

Figure~15 shows the fudge factor $f(E)$ corresponding to the specific transformation $E\to E'$
with $\Delta m^2_{ee}= \Delta m^{2\,\prime}_{ee}$, for both NH and IH. The factor diverges at threshold
but, for neutrino energies sufficiently above $E_T$ (say, $E\gtrsim 2.1$~MeV) it takes  
values of $O(10\%)$, and can become as low as O(2\%) in the high-energy part of the spectrum.
Shape variations of about this size may still be tolerated within current uncertainties on the reactor 
spectrum \cite{Hube,Spec}, and thus should be kept under control in future JUNO-like
experiments.

\begin{figure}[b]
\vspace*{0.0cm}
\hspace*{.0cm}
\includegraphics[scale=0.33]{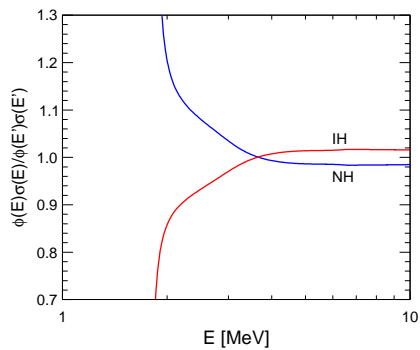}
\vspace*{-0.0cm}
\caption{\label{f15} Energy profiles of the fudge factors which would ``undo'' the reactor spectral changes
induced by the changes $E\to E'$ reported in Fig.~11.}
\end{figure}

If one applies, at the same time, the transformation $E\to E'$ in Fig.~11 and the fudge factor
$f(E)$ in Fig.~15 to the case of true NH, the results in Fig.~12 remain basically the same (i.e., 
the ``wrong'' IH is preferred), but the spectral mismatch around threshold in Fig.~13 is
largely cured, and the total $\chi^2$ in Fig.~14 drops from $\sim 360$ to $\sim 22$. In this case,
the best-fit IH spectrum is almost completely degenerate with the true NH spectrum. 

Fig.~16
shows the corresponding $\chi^2$ density, which is now dominated by the residual shape mismatch 
of the geoneutrino energy spectrum, whose step-like features still occur ``at the wrong energy''
and slightly break the degeneracy. Geoneutrinos thus offer an additional handle to self-calibrate 
the low-energy scale to some extent, as also pointed out in another context \cite{Kopp}.  

Summarizing, an energy scale transformation as in Fig.~11 is able to swap the 
hierarchy in the fit (Fig.~12), but it also induces a mismatch in the spectral features around threshold (Fig.~13) and thus
a very high $\chi^2$ value at best fit (Fig.~14), which could be used as a diagnostic, self-calibration
tool. However, specific variations of the reactor spectrum shape (Fig.~15) can largely ``undo'' the
low-energy mismatch, leaving only a residual misfit in the geoneutrino spectral shapes (Fig.~16)
which could be used a secondary self-calibration tool. There is thus a subtle interplay between 
energy scale systematics and spectral shape uncertainties, which need to be kept under control
in order to discriminate the hierarchy and to get unbiased estimates of the $\nu_e$ oscillation 
parameters.

\begin{figure}[t]
\vspace*{0.0cm}
\hspace*{.0cm}
\includegraphics[scale=0.36]{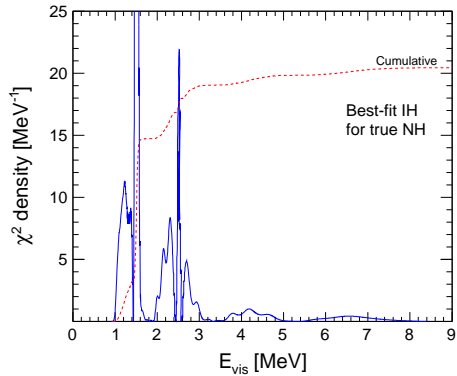}
\vspace*{-0.0cm}
\caption{\label{f16} $\chi^2$ density and its cumulative distribution at best fit,
after modifying the shape of the reactor energy spectrum as in Fig.~15. To be compared
with Fig.~14; see the text for details.}
\end{figure}

\subsection{Energy scale transformation $E\to E'$ with $E=E'$ at threshold}

In the previous section, it has been shown that the specific
choice $\Delta m^2_{ee}= \Delta m^{2\,\prime}_{ee}$ in Eq.~(\ref{iter}) leads to both energy scale and spectral 
variations mainly localized at relatively low energies, $E_\mathrm{vis}\lesssim 3$~MeV. However,
other choices in Eq.~(\ref{iter}) may move the relevant variations to 
the high-energy part of the spectrum. In particular, one may choose the ratio $ \Delta m^{2\,\prime}_{ee}/\Delta m^2_{ee} $
in Eq.~(\ref{iter}) so as to get $E/E'=1$ just at threshold ($E=E_T$). For the considered JUNO set up, 
this choice corresponds to take $  \Delta m^{2\,\prime}_{ee}/\Delta m^2_{ee} \simeq 1.022$ (0.978) for true
NH (IH). Figure~17 shows the corresponding  profile of the energy ratio $E/E'$, 
which is $\lesssim 2$ permill for $E\lesssim 3$~MeV, but grows up to $\sim 2\%$ at higher energies.

\begin{figure}[b]
\vspace*{.0cm}
\hspace*{.0cm}
\includegraphics[scale=0.32]{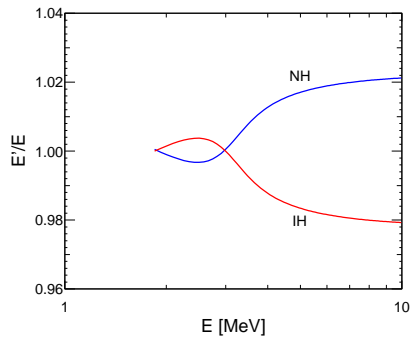}
\vspace*{-0.cm}
\caption{\label{f17} Profile of the neutrino energy ratio $E'/E$ which flips the sign of the
hierarchy-dependent phase $\varphi$ in the JUNO experiment, for the case  $  \Delta m^{2\,\prime}_{ee}/\Delta m^2_{ee} \simeq 1.022$ (0.978) in 
NH (IH). The profiles are shown for $E\geq E_T=1.806$~MeV.}
\end{figure}

If we fit the prospective JUNO data (for true NH) with $E\to E'$ as in Fig.~17, the preferred
value of $\alpha$ is shifted down to $\sim -1$ 
(similarly to Fig.~12), and the best-fit value of $\Delta m^2_{ee}/(10^{-3}\ \mathrm{eV}^2)$ is 
shifted to $\sim 2.48 \simeq 1.022\times 2.43$ as expected (not shown).  In this case, the effect of the energy scale
variation in Fig.~17 is quite dramatic: at best fit, not only the wrong hierarchy is preferred, 
but the value of $\Delta m^2_{ee}$ is biased by
an order of magnitude more than its prospective $1\sigma$ accuracy (as reported in Table~I). However, also in this
case the degeneracy between the spectra for
true and wrong hierarchy is not complete: the best fit in the ``wrong'' IH is very bad ($\chi^2\simeq 280$), and
receives contributions mainly from the high-energy part of the spectrum, $E_\mathrm{vis}\gtrsim 3$~MeV (not shown). 

A misfit of the high-energy tail of the spectrum might be used as a diagnostic of systematic energy scale deviations in 
that region; however, the misfit could be largely compensated by appropriate variations of the reactor spectrum shape 
via a fudge factor $f(E)$, analogously to the case discussed in the previous section.  Figure~18 shows the fudge
factor which brings the best-fit spectrum for the wrong hierarchy in much closer agreement with the original
spectrum for the true hierarchy (either normal or inverted), for energy variations as in Fig.~17. If deviations as large
as in Fig.~18 are allowed within reactor spectral shape uncertainties \cite{Hube,Spec}, then the degeneracy 
between true and wrong hierarchy would be almost complete along the whole energy range, with
residual misfits located mainly in the geoneutrino energy region. Indeed, by applying the fudge factor in
Fig.~18, the ``wrong'' hierarchy is still preferred, but the $\chi^2$ at best fit drops by an order of magnitude, and 
the best-fit value of $\Delta m^2_{ee}$ is brought back to the original value $\sim 2.43\times 10^{-3}$~eV$^2$
(not shown).

\begin{figure}[t]
\vspace*{0.0cm}
\hspace*{.0cm}
\includegraphics[scale=0.33]{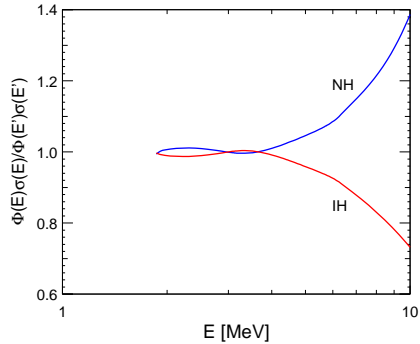}
\vspace*{-0.0cm}
\caption{\label{f18} Energy profiles of the fudge factors which would ``undo'' the reactor spectral changes
induced by the changes $E\to E'$ reported in Fig.~17.}
\end{figure}

\subsection{Further comments on energy scale and spectral shape deviations}

The simultaneous occurrence of energy scale deviations $E\to E'$ as in Eq.~(\ref{iter}) and of
reactor spectral shape deviations $f(E)$ as in Eq.~(\ref{fudge}) makes true and wrong hierarchies 
nearly degenerate across the whole visible energy range, with
small residual misfits mainly located in the geoneutrino energy region. If such deviations
are allowed within the systematic uncertainties of the JUNO experiment, the discrimination 
of the hierarchy would be seriously compromised, being degraded at significance level necessarily lower than 
the $\sim 2\sigma$ estimated in Sec.~VI.   

At present it is premature ---if not impossible--- to guess the final accuracy on the
energy scale achievable with dedicated calibration experiment and detector simulations, 
as well as the reduction of spectral shape uncertainties reachable after
 the current campaign of high-statistics, near-detector measurements. 
It is also not particularly useful to embed such deviations by means of
arbitrary functional forms (e.g., polynomials) and corresponding pulls in the fit,
unless such functions can also cover the family of nonlinear deviations implied by 
Eqs.~(\ref{iter}) and (\ref{fudge}) [see also \cite{Ciuf,Snow}]. We are thus
approaching a new and unusual situation in neutrino physics, which was already 
highlighted in another context \cite{RCCN}: spectral measurements with very high statistics require
dedicated studies of the ``shape'' of nonlinear systematics, which are
not necessarily captured by simply adding a few more pulls and penalties in the fit.

Finally, we emphasize that, in our approach with free $\alpha$,
the discrimination of the hierarchy is already
compromised when $\alpha=\pm 1$ is misfitted as $\alpha=0$ (case of ``undecidable'' hierarchy) rather than
as $\alpha=\mp 1$ (case of ``wrong'' hierarchy).  In particular, the energy scale deviation which would 
bring $\alpha=\pm 1$ to $\alpha=0$ (``canceling'' 
the hierarchy-dependent oscillation phase $\varphi$) obeys the equation:
\begin{equation}
\frac{\Delta m^2_{ee}\,L}{2E}\pm \varphi(E) = \frac{\Delta m^{2\,\prime}_{ee} \,L}{2E'}
\ ,
\label{Escale2}
\end{equation}
which is approximately solved by 
\begin{equation}
\frac{E'}{E}\simeq \frac{\Delta m^{2\prime}_{ee}} {\Delta m^2_{ee}}\mp s^2_{12}\frac{\delta m^2}{\Delta m^2_{ee}}
\left(1-\frac{\sin\delta(E)}{2\delta(E)\sqrt{P^{2\nu}_\mathrm{vac}(E)}}\right)\ .
\label{iter2}
\end{equation}
In comparison with Eq.~(\ref{iter}), the nonlinear term in the above equation is a factor of two smaller. 
Therefore, energy scale deviations of about half the size discussed in the previous
two Sections (as well as spectral deviations $f(E)$ reduced by a similar factor)
are already sufficient to compromise the hierarchy determination, bringing 
the fit close to the null case $\alpha=0$, as we have explicitly verified numerically in various cases.
For instance, Fig.~19 shows the fit results assuming $\Delta m^{2\prime}_{ee} = \Delta m^2_{ee}$ in Eq.~(\ref{iter2}), 
In this case, the deviations of $E'/E$ from unity are exactly 1/2 smaller than in Fig.~11, and the true value $\alpha=1$
is misfitted as $\alpha\simeq -0.2\pm 0.5$ at $\pm1\sigma$, which is more consistent with $\alpha=0$ (undecidable hierarchy)
than with $\alpha=\pm 1$ (true or wrong hierarchy). In this respect, it is useful to compare Fig.~19  
with both Fig.~7 (best fit 
at the true hierarchy) and Fig.~12 (best fit at the wrong hierarchy).

In a sense, the challenge of the energy scale may actually be 
greater than pointed out in \cite{Voge} and \cite{Snow,Ciuf}), by approximately a factor of two. 
In particular, in order to reject cases leading to $\alpha\simeq 0$ with, say, $3\sigma$ confidence, the ratio $E'/E$ should be kept
close to unity at the few permill level over the whole reactor neutrino energy range. Moreover, as shown
in this section, the conspiracy of energy scale and spectrum shape systematics may lead to an even stronger degeneracy 
between cases with different values of $\alpha$ (e.g., $\alpha=\pm 1$ versus $\alpha=\mp1$, or versus $\alpha=0$), making it
very difficult to prove the occurrence of a non-$L/E$ oscillation phase $\varphi$ with a definite sign.  
An additional detector close to the main reactors might help to mitigate the impact of such systematics (see \cite{Snow,Ciuf}), provided 
that the near and far detector responses are proven to be very similar at all energies, 
so as to cancel out correlated uncertainties on both the
$x$ and $y$ axes of event spectra. All these delicate issues definitely require further investigations, in order
to prove the feasibility of hierarchy discrimination at reactors with sufficient statistical significance.
   
\begin{figure}[t]
\vspace*{0.0cm}
\hspace*{.0cm}
\includegraphics[scale=0.33]{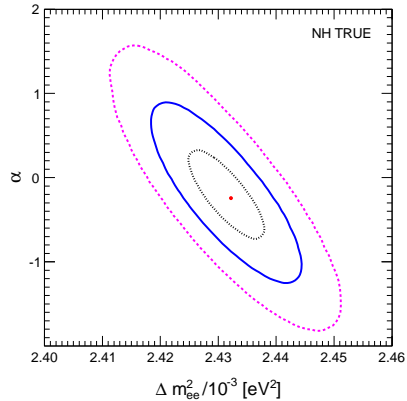}
\vspace*{-0.0cm}
\caption{\label{f19} Constraints in the plane $(\Delta m^2_{ee},\,\alpha)$ for true NH, after applying 
the energy scale variations of Fig.~11 reduced by a factor of two. The best fit is roughly half-way between the true hierarchy
$(\alpha=+1)$ and the wrong
hierarchy ($\alpha=-1$). See the text for details.}
\end{figure}

\section{Summary and conclusions}

Medium-baseline reactor neutrino experiments can offer unprecedented opportunities to
probe, at the same time, all the parameters which govern the mixing of $\overline\nu_e$ with the neutrino mass states, namely, the 
mixing angles  $\theta_{12}$ and $\theta_{13}$, the two squared mass differences $\delta m^2$ and $\Delta m^2$, and the hierarchy
parameter $\alpha$ ($+1$ for NH and $-1$ for IH). These goals largely justify the currents efforts towards
the construction of such experiments, as currently envisaged by the JUNO and RENO-50 projects.

In this context,
 we have revisited some issues raised by the need of precision calculations
and refined statistical analyses of reactor event spectra. In particular, we have shown  
how to include analytically IBD recoil effects in binned and unbinned spectra, 
via appropriate modifications of the energy resolution function (Sec.~III). We have also generalized the oscillation probability
formula by including analytically matter propagation and multiple reactor damping effects, and by treating the
parameter $\alpha$ as a continuous --- rather than discrete --- variable (Sec.~IV). The determination 
of the hierarchy is then transformed from a test of hypothesis to 
a parameter estimation, with a sensitivity given by the statistical distance of the true case 
(either $\alpha=+1$ or $\alpha=-1$) from the ``undecidable'' case ($\alpha=0$). 
Numerical experiments have been performed for the specific experimental set up envisaged for 
the JUNO experiment, assuming a realistic sample of $O(10^5)$ medium-baseline reactor events, 
plus geoneutrino and far-reactor backgrounds, via an unbinned $\chi^2$ analysis. 
We have found a typical sensitivity to the hierarchy
slightly below $2\sigma$ in JUNO, and significant prospective improvement upon current
errors on the oscillation parameters (see Table~I and Figs.~7--9), as far as systematic uncertainties are limited 
to reactor and geoneutrino normalization errors (Sec.~V). 

Further systematic uncertainties, associated to energy scale and spectrum shape distortions, may seriously compromise 
the hierarchy sensitivity and may also bias the oscillation parameters (Sec.~VI). In 
particular,  specific energy scale variations --- for which we have provided compact
expressions --- can move the reconstructed value of $\alpha$ away from the true one; e.g.,
$\alpha=+1$ can be misfitted as $\alpha\simeq -1$ or as $\alpha\simeq 0$. However, the
overall fit is generally very bad in such cases, since the reactor spectrum is also distorted at either low or high energies 
with respect to expectations. In principle, these shape distortions might be used as a diagnostic
of energy scale errors; however, they might also be compensated by opposite
ones within current shape uncertainties, in which case the degeneracy between ``true'' and ``wrong''
values of $\alpha$ would be almost complete (up to residual, unbalanced distortions of geoneutrino spectra).
For instance, the joint occurrence of  distortions as in Figs.~11 and 15 (or as in Figs.~17 and 18) would
essentially flip the hierarchy parameter ($\alpha=\pm 1\to \alpha\simeq \mp1$) with only a modest increase in the $\chi^2$.
Distortions of about half this size are sufficient
to bring the fit close to the case of ``undecidable hierarchy'' as in Fig.~19 ($\alpha=\pm 1\to \alpha\simeq 0$), 
thus compromising the hierarchy discrimination.
It is thus very important to control, at the same time, the systematic uncertainties 
on both the $x$-axis (energy scale) and the 
$y$-axis (spectrum shape) of measured and simulated reactor event spectra,
with an accuracy sufficient to reject the above distortions at high confidence level.


\acknowledgments

{\em Acknowledgments.} 
This work is supported by the Italian Istituto Nazionale di Fisica 
Nucleare (INFN) through the ``Theoretical Astroparticle Physics''  project.


\end{document}